\documentclass[pra,aps,twocolumn,nopacs,superscriptaddress,nofootinbib]{revtex4}

\usepackage{graphicx}
\usepackage{dcolumn}
\usepackage{bm}
\usepackage{bbm}
\usepackage{amsmath}
\usepackage{epsfig}
\usepackage{indentfirst}
\usepackage{psfrag}
\usepackage{subfigure}
\usepackage{amssymb}
\usepackage{color}
\usepackage[colorlinks,linkcolor=blue,citecolor=blue,urlcolor=blue,hyperindex,driverfallback=dvipdfm]{hyperref}
\usepackage[T1]{fontenc}

\def\ii{{\rm i}}  \def\ee{{\rm e}}
\def\rb{{\bf r}}  \def\Rb{{\bf R}}    \def\vb{{\bf v}}
    \def\zz{\hat{\bf z}}  \def\nn{\hat{\bf n}}
\def\kb{{\bf k}}    
    
\def\Eb{{\bf E}}

      \def\Pb{{\bf P}}

\def\sb{{\bf s}}

\newcommand{\abs}[1]{\vert #1 \vert}

\begin{document}
%\title{Nonlinear nanophotonics with electron beams\\
%Nanoscale nonlinear spectroscopy with ultrafast electron miscroscopy ... or PINEM}
\title{Nanoscale Nonlinear Spectroscopy with Electron Beams}
\author{Andrea~Kone\v{c}n\'{a}}
\affiliation{ICFO-Institut de Ciencies Fotoniques, The Barcelona Institute of Science and Technology, 08860 Castelldefels (Barcelona), Spain}
\author{Valerio~Di~Giulio}
\affiliation{ICFO-Institut de Ciencies Fotoniques, The Barcelona Institute of Science and Technology, 08860 Castelldefels (Barcelona), Spain}
\author{Vahagn~Mkhitaryan}
\affiliation{ICFO-Institut de Ciencies Fotoniques, The Barcelona Institute of Science and Technology, 08860 Castelldefels (Barcelona), Spain}
\author{Claus~Ropers}
\affiliation{4th Physical Institute-Solids and Nanostructures, University of Gottingen, G\"ottingen 37077, Germany}
\author{F.~Javier~Garc\'{\i}a~de~Abajo}
\email[Corresponding author: ]{\\ javier.garciadeabajo@nanophotonics.es}
\affiliation{ICFO-Institut de Ciencies Fotoniques, The Barcelona Institute of Science and Technology, 08860 Castelldefels (Barcelona), Spain}
\affiliation{ICREA-Instituci\'o Catalana de Recerca i Estudis Avan\c{c}ats, Passeig Llu\'{\i}s Companys 23, 08010 Barcelona, Spain}

\begin{abstract}
We theoretically demonstrate the ability of electron beams to probe the nonlinear photonic response with nanometer spatial resolution, well beyond the capabilities of existing optical techniques. Although the interaction of electron beams with photonic modes is generally weak, the use of optical pumping produces stimulated electron-light interactions that can reach order-unity probabilities in photon-induded near field electron microscopy (PINEM). Here, we demonstrate that PINEM can locally and quantitatively probe the nonlinear optical response. Specifically, we predict a dependence of PINEM electron spectra on the sample nonlinearity that can reveal the second-harmonic (SH) response of optical materials with nanometer resolution, observed through asymmetries between electron energy losses and gains. We illustrate this concept by showing that PINEM spectra are sensitive to the SH near field of centrosymmetric structures and by finding substantial spectral asymmetries in geometries for which the linear interaction is reduced.
\end{abstract}
\date{\today}
%\pacs{79.20.Uv,73.20.Mf,42.50.Ct,78.67.-n}
\maketitle

% 79.20.Uv Electron energy loss spectroscopy
% 73.20.Mf Collective excitations
% 42.50.Ct Quantum description of interaction of light and matter; related experiments
% 78.67.−n Optical properties of low-dimensional, mesoscopic, and nanoscale materials and structures

\section{Introduction} % ---------------------

Electron microscope spectroscopies have evolved into a powerful set of techniques capable of providing structural and dynamical information of materials with nanometer/femtosecond/meV space/time/energy resolution \cite{E96,BFZ09,paper151,PLZ10,KLD14,KGK14,FES15,PLQ15,EFS16,RB16,VFZ16,KML17,FBR17,LTH17,PRY17,paper306,paper311,paper332,WDS19,KLS19,DNS19,HHP19,paper338}. In particular, low-loss electron energy-loss spectroscopy (EELS) can nowadays access local spectral information on plasmons in metallic nanostructures \cite{paper149,RB13,KS14,HTH15,GBL17,KGS18}, excitons in semiconductors \cite{TLM15}, phonons in ionic crystals \cite{KLD14,LTH17} and graphene \cite{SSB19}, and atomic vibrations in molecules \cite{RAM16,HHP19}. Additionally, ultrafast temporal resolution is achieved in PINEM by synchronizing the time of arrival of femtosecond electron and optical pulses at the sample \cite{BFZ09,paper151,PLZ10,KGK14,FES15,PLQ15,EFS16,RB16,VFZ16,KML17,PRY17,paper306,paper311,paper332,KLS19,DNS19,WDS19}. Recent proposals further extend light-matter interactions in PINEM to produce attosecond electron pulses \cite{PRY17,MB17}, electron entanglement \cite{K19}, probe photon statistics \cite{paper3xy}, and perform quantum computations \cite{RML19}.

The high spatial resolution enabled by electron beams could also find application in the mapping of the nonlinear optical response in nanostructures, which is important from both fundamental and applied viewpoints to better understand and improve the performance of nonlinear nanophotonic devices \cite{KZ12,PSL18,BBM15}. However, despite the widespread use of electron-beam spectroscopies to characterize the linear response of nanomaterials, the higher-order nonlinear response is generally considered unreachable because of the weak interaction between individual beamed electrons and sample excitations. This scenario is substantially changed in PINEM, where sample modes are populated through external optical pumping to high occupation numbers that can yield scattering probabilities of order unity, effectively resulting in multiple quanta exchanges between the electron probe and the optical field, observed to generate up to hundreds of loss and gain orders \cite{KLS19,DNS19}. Additionally, the femtosecond duration of both electron and optical pulses allows employing high light intensities that can trigger substantial nonlinearities without damaging the sample. The prospects are therefore excellent for the use of PINEM to probe the nonlinear optical response of materials at length scales determined by the subnanometer transversal size of focused electron beams. This represents a radical improvement in terms of noninvasiveness, intrinsic phase sensitivity, and spatial resolution compared to existing nonlinear characterization techniques relying on either far- \cite{BPS98,B08_3} or near-field optics \cite{BPS98,ZS00,BBH03,ZCB08,NVF09,MHG17}.
% and are limited by either diffraction or the physical size of the probing tip to tens of nanometers at best.

In this Letter, we theoretically demonstrate the potential of PINEM to quantitatively probe the nonlinear optical response with nanometer spatial resolution. Specifically, we focus on the sampling of SH fields, which are revealed as asymmetries in the PINEM spectra. We illustrate this concept by first considering spherical gold nanoparticles, which, despite their centrosymmetry, display an evanescent SH near field that gives rise to substantially modified transmission electron spectra under attainable ultrafast illumination intensities below the damage threshold. We further explore the interaction with nanorods as an example of configuration in which linear-field coupling is strongly reduced, further increasing the spectral asymmetry to the 10\% level. Our results support PINEM performed with variable illumination frequency as a nonlinear optical characterization technique with unsurpassed combination of spatial and spectral resolution.

% FIGURE 1 -----------------------------------------
\begin{figure}
\centering
\includegraphics[width=0.42\textwidth]{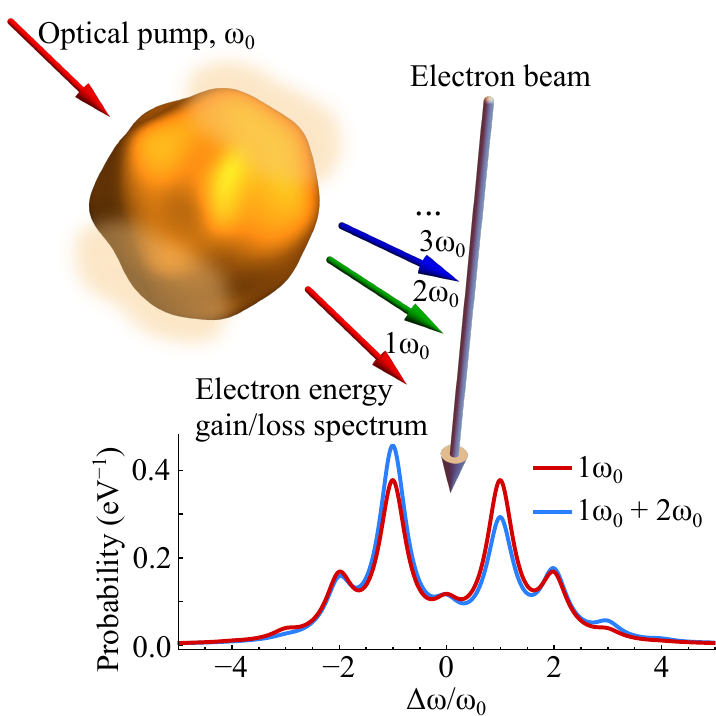}
\caption{Nanoscale sampling of the nonlinear optical response. An optical pump at the sampling frequency $\omega_0$ illuminates a nanoparticle in coincidence with the arrival of a focused electron. If the particle responds linearly, the EELS spectrum presents a symmetric distribution of stimulated loss ($\omega>0$) and gain ($\omega<0$) features (red curve). Nonlinear response in the particle generally produces harmonic optical fields at multiples of $\omega_0$, which are revealed through an asymmetric EELS spectrum (blue curve). We illustrate this effect by taking $\abs{\beta_1}=1$, $\abs{\beta_2}=0.1$, $\delta=0$, $\hbar\omega_0=1$~eV, and a Lorentzian broadening of 300~meV (see main text).} 
\label{Fig1}
\end{figure}

Energy-momentum mismatch prevents absorption or emission of photons by electrons in free space. In contrast, translational-symmetry breaking in illuminated nanostructures enables such coupling \cite{H99,paper114}, which is mediated by near-field components that give rise to multiple exchanges of photons between the electron and the optical field (Fig.\ \ref{Fig1}). More precisely, when neglecting nonlinear optical fields, the electron-light interaction is fully captured by the parameter \cite{paper151,PLZ10,paper272} $\beta_1=(e/\hbar\omega_0)\int dz\,E_z^{(1)}\ee^{-\ii\omega_0 z/v}$, where $E_z^{(1)}$ is the linear electric field component along the direction of the electron velocity $\vb=v\zz$, integrated over positions $z$ along the electron trajectory, and $\omega_0$ is the light frequency. The transmitted electron spectrum is then characterized by loss ($\ell<0$) and gain ($\ell>0$) peaks (electron energy change $\ell\hbar\omega_0$) of integrated probability $P_\ell=J_\ell^2(2|\beta_1|)$ defined in terms of Bessel functions $J_\ell$ (Fig.\ \ref{Fig1}, red curve). We remark that, although multiple peaks are produced in the spectrum, the interaction is fully controlled by the single parameter $\beta_1$, which is linear in the electric field.

As we show below, the nonlinear response associated with the nanostructure can produce near fields at frequencies that are multiples of $\omega_0$ and result in asymmetries of the electron spectrum (Fig.\ \ref{Fig1}, blue curve) like those observed under external illumination consisting of superimposed harmonics \cite{PRY17}. In what follows, we focus on gold nanoparticles, in which the bulk second-order nonlinear response cancels due to inversion symmetry of the crystal lattice, while the surface SH response is relatively large compared with other materials \cite{BCJ1968,SMW1974,SSF1980,GSW18,B08_3} and can be substantially enhanced due to field amplification mediated by surface plasmons \cite{KZ12,PSL18,BBM15}. For simplicity, we neglect higher-order nonlinear terms, which should be comparatively smaller under the conditions considered below.

\section{Theoretical description of nonlinear PINEM}

We extend previously developed PINEM theory \cite{paper151,PLZ10,paper272} to incorporate both the fundamental and SH fields in the electron-light interaction. While previous works have considered superimposing fundamental and higher-harmonic fields in PINEM \cite{PRY17}, we now take into account the intrinsic SH response generated in the nanostructure. More precisely, we consider an incident electron with small energy and momentum spread relative to central values $E_0$ and $\hbar\kb_0$, so that its wave function can be written as $\psi(\rb,t)=\ee^{\ii(\kb_0\cdot\rb-E_0t/\hbar)}\phi_0(\rb,t)$ in terms of a smooth function $\phi_0(\rb,t)$ that undergoes only small variations over each optical period. After PINEM interaction, the transmitted electron wave function is given by this expression with $\phi_0$ replaced by $\phi=\phi_0\sum_\ell f_\ell\,\ee^{\ii\ell\omega_0(z/v-t)}$, where the electron is taken to move along $z$ and the sum extends over components associated with an effective number of exchanged photons $\ell$ ($>0$ for gain and $<0$ for loss). The amplitudes of these components are found to be (see Appendix)
\begin{align}
f_\ell &=\ee^{\ii\ell\arg\{-\beta_1\}} \!\!\! \sum_{n=-\infty}^{\infty} \!\!\! \ee^{-\ii n\delta} J_{\ell+2n}\left( 2|\beta_1|\right) J_n\left( 2|\beta_2|\right),
\nonumber
\end{align}
where
\begin{align}
\beta_j=\frac{e}{\hbar j\omega_0 }\int_{-\infty}^{\infty} dz\,E_{\textrm{z}}^{(j)}(\Rb_0,z)\ee^{-\ii j\omega_0z/v}
\label{Eq:beta}
\end{align}
describes the interaction with the fundamental ($j=1$) and SH ($j=2$) fields of frequency $j\omega_0$,
\begin{align}
\delta=\arg\{\beta_2\} - 2\arg\{\beta_1\}
\label{Eq:delta}
\end{align}
captures the dependence on the relative phase of both $j$ fields, and the impact parameter $\Rb_0=(x_0,y_0)$ defines the position of the electron beam under the assumption that its transversal size is small compared with the optical fields under consideration. The probability associated with an electron energy change $\ell\hbar\omega_0$ is simply given by $P_\ell=|f_\ell|^2$, which obviously depends on the coupling strengths $|\beta_j|$, but also on the phase difference $\delta$. Incidentally, Eqs.\ (\ref{Eq:beta}) and (\ref{Eq:delta}) predict a phase $\delta$ independent of any displacement in the position $z$ of the field relative to the electron wave function.

% FIGURE 2 -----------------------------------------
\begin{figure*}
\centering
\includegraphics[width=0.95\textwidth]{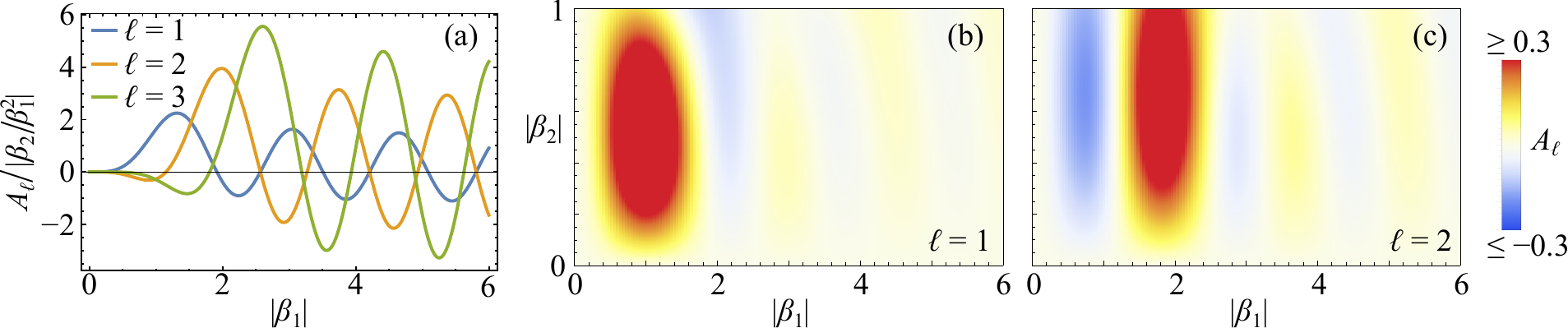}
\caption{Second-harmonic-induced asymmetry in the PINEM spectra. We plot the asymmetry parameter $A_\ell$ [Eq.\ (\ref{Al})] in the $|\beta_2|\ll1$ limit (a) and for larger values of this parameter (b). We consider selected sideband orders $\ell$ and take $\delta=0$.}
\label{Fig2}
\end{figure*}

In practice, we expect to deal with small values of the SH coupling coefficient $|\beta_2|$, for which the probability of the sideband $\ell$ reduces to
\begin{align}
P_\ell=J_\ell^2(2|\beta_1|)+|\beta_2|\;C_\ell(2|\beta_1|)\cos\delta,
\label{Plsmallx2}
\end{align}
where $C_\ell(x)=2J_\ell(x)\left[J_{\ell+2}(x)-J_{\ell-2}(x)\right]$ (see Appendix). This expression shows that $P_\ell$ deviates maximally from the linear PINEM regime when $\delta$ is a multiple of $\pi$, a result that is also maintained for arbitrarily large values of $|\beta_2|$ (see Appendix). Importantly, SH components enter the PINEM probability through a linear correction in the SH field amplitude instead of its intensity, thus facilitating the determination of the nonlinear material response for the expected low values of $|\beta_2|$. Additionally, when the linear PINEM coefficient vanishes ($\beta_1=0$), one obtains a regular PINEM spectrum with sidebands separated by $2\hbar\omega_0$, as determined by the SH coupling coefficient $\beta_2$, which for $|\beta_2|\ll1$ produces probabilities $P_\ell\approx|\beta_2|^{2\ell}/\ell!^2$.

As a way of capturing the spectral asymmetry observed in Fig.\ \ref{Fig1} due to nonlinear interactions, we define the parameter
\begin{align}
A_\ell=P_\ell-P_{-\ell}
\label{Al}
\end{align}
(difference between gain and loss probabilities in sidebands $\ell$ and $-\ell$), which for small $|\beta_2|$, using Eq.\ (\ref{Plsmallx2}), becomes $A_\ell\propto|\beta_2/\beta_1^2|$, with a coefficient of proportionality $2|\beta_1|^2C_\ell(|\beta_1|)\cos\delta$ that depends on the illumination intensity. Obviously, the ratio $|\beta_2/\beta_1^2|$ is independent of light intensity, therefore facilitating the determination of the nonlinear SH response upon direct inspection of the asymmetry parameters $A_\ell$. Additionally, the order $\ell$ that is best suited to resolve the nonlinear behavior depends on the range of $|\beta_1|$, as shown in Fig.\ \ref{Fig2}(a) for small $|\beta_2|$ and Fig.\ \ref{Fig2}(b,c) for a larger range of this parameter.

In what follows, we calculate the SH field by considering a distribution of surface dipoles oriented along the local surface normal $\nn_\sb$ with a polarizability per unit area at each surface position $\sb$ given by \cite{BBR10_2}
\begin{equation}
\Pb_\sb^{(2)}=\chi_{\perp\perp\perp}\left[\nn_\sb\cdot\Eb^{(1)}(\sb)\right]^2\nn_\sb,
\label{P2}
\end{equation}
where $\chi_{\perp\perp\perp}$ is the dominant component of the SH surface susceptibility (other tensor components are negligible in metals \cite{KTR04, WRA09, BBR10_2, TYK18}), and the linear field $\Eb^{(1)}$ needs to be evaluated for the illumination frequency $\omega_0$ at a point immediately inside the metal. We obtain $\Eb^{(1)}$ by solving Maxwell's equations with a light plane wave as a source and the gold described through its tabulated frequency-dependent dielectric function \cite{JC1972}. From here, we obtain the SH field $\Eb^{(2)}$ by again solving those equations with the surface dipole distribution [Eq.\ (\ref{P2})] as a source. These fields are then inserted into Eq.\ (\ref{Eq:beta}) to produce the coupling parameters $\beta_j$. Incidentally, the SH field entering $\beta_2$ is equivalently obtained using the reciprocity theorem from the field produced by the passing electron at frequency $2\omega_0$ on the particle surface, which results in a substantial reduction of computation time (see Appendix).

% FIGURE 3 -----------------------------------------
\begin{figure}
\centering
\includegraphics[width=0.46\textwidth]{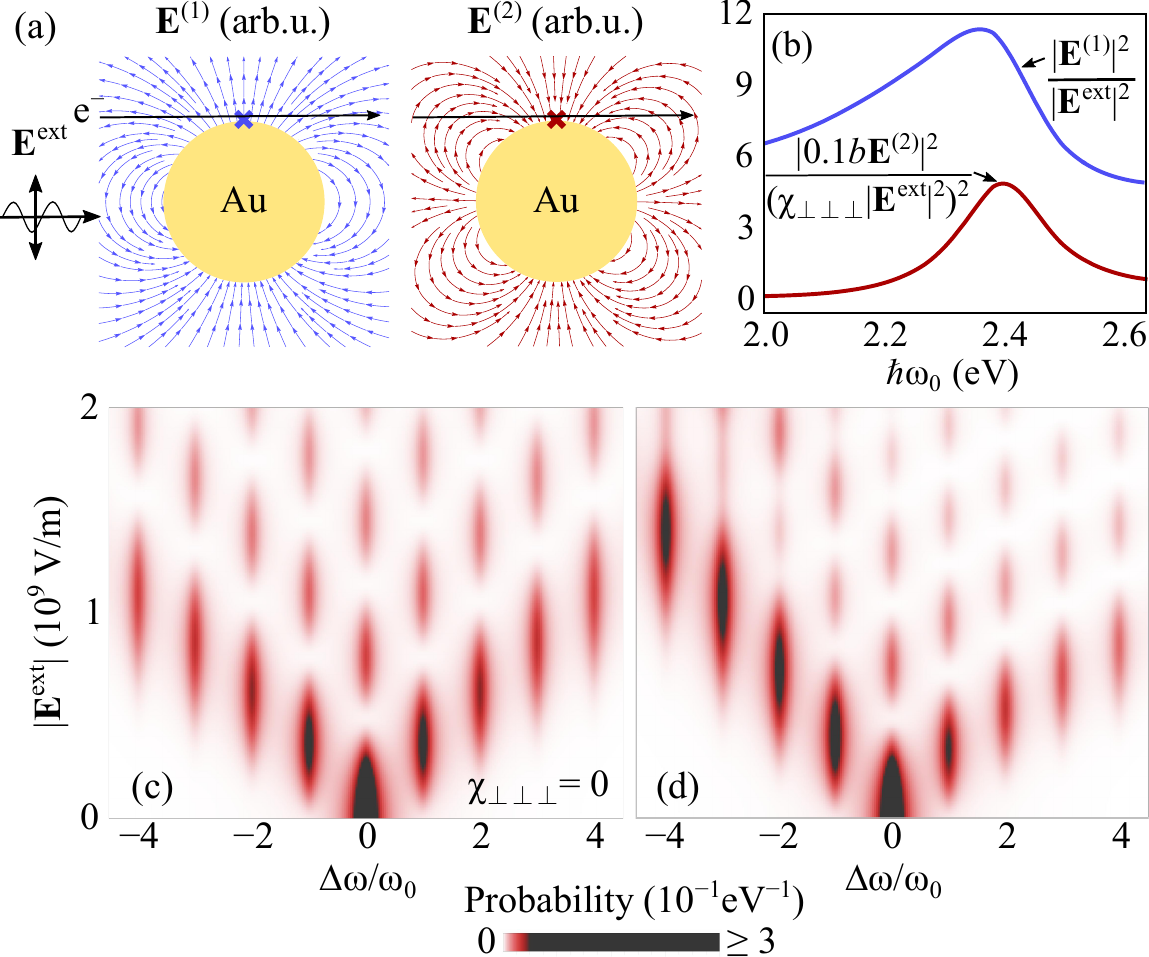}
\caption{PINEM sampling of the second-harmonic near field in a gold sphere. (a) A linearly polarized light plane wave (external field $\Eb^\mathrm{ext}$, frequency $\omega_0$) excites a gold sphere (20\,nm diameteter) giving rise to induced fields $\Eb^{(1)}$ and $\Eb^{(2)}$ at the fundamental ($\omega_0$) and SH ($2\omega_0$) frequencies, respectively. We show electric field lines outside the particle in the plane defined by the light polarization and propagation directions for $\hbar\omega_0=2.4$~eV. (b) Spectral dependence of the linear (blue curve) and SH (red curve) near-field intensity at the position of the crosses (distance $b=2\,$nm from the surface), normalized by using $|\Eb^\mathrm{ext}|$, the SH susceptibility $\chi_{\perp\perp\perp}$, and the distance $b$. (c) Spectral and light-field-amplitude dependence of the PINEM electron probability for $\hbar\omega_0=2.4$~eV in the absence of nonlinear particle response ($\chi_{\perp\perp\perp}=0$) for 100\,keV electrons passing 2\,nm outside the surface along the trajectory indicated by the black arrows in (a). We introduce a 0.3\,eV Lorentzian broadening for clarity. (d) Same as (c), but including SH fields for a typical nonlinear susceptibility \cite{KTR04,WRA09} $\chi_{\perp\perp\perp}=10^{-18}~\mathrm{m^2/V}$.} 
\label{Fig3}
\end{figure}

\section{Probing the second-harmonic near field in centrosymmetric structures}

Although inversion symmetry prevents far-field SH generation, an evanescent field at frequency $2\omega_0$ can still exist in the vicinity of such illuminated nanostructures and interact with a passing electron to produce PINEM asymmetries. We illustrate this possibility by considering a spherical gold nanoparticle (Fig.\ \ref{Fig3}) based on an analytical solution of this problem in the quasistatic limit (see details in Appendix), which, given the small diameter of the particle under consideration (20\,nm), we find to be in excellent agreement with numerically obtained retarded calculations. As expected \cite{DSE1999,DSH04,RBB07}, the linear near field exhibits a characteristic dipolar pattern oriented along the incident polarization, while the SH field displays a quadrupolar profile [Fig.\ \ref{Fig3}(a)]. Additionally, a prominent $\sim2.4$~eV particle plasmon is observed in the spectral dependence of both linear and SH near fields [Fig.\ \ref{Fig3}(b)], with maximum intensity at the sphere poles. For a 100\,keV electron passing 2\,nm away from the upper pole, we obtain a regular PINEM spectral profile describable through the probabilities $J_\ell^2(2|\beta_1|)$ when neglecting nonlinear effects [Fig.\ \ref{Fig3}(c)], while inclusion of SH response produces a substantial asymmetry for experimentally feasible light field amplitudes [Fig.\ \ref{Fig3}(d)], thus corroborating that the electron can indeed sample SH near fields despite the symmetry of the particle.

% FIGURE 4 -----------------------------------------
\begin{figure}
\centering
\includegraphics[width=0.48\textwidth]{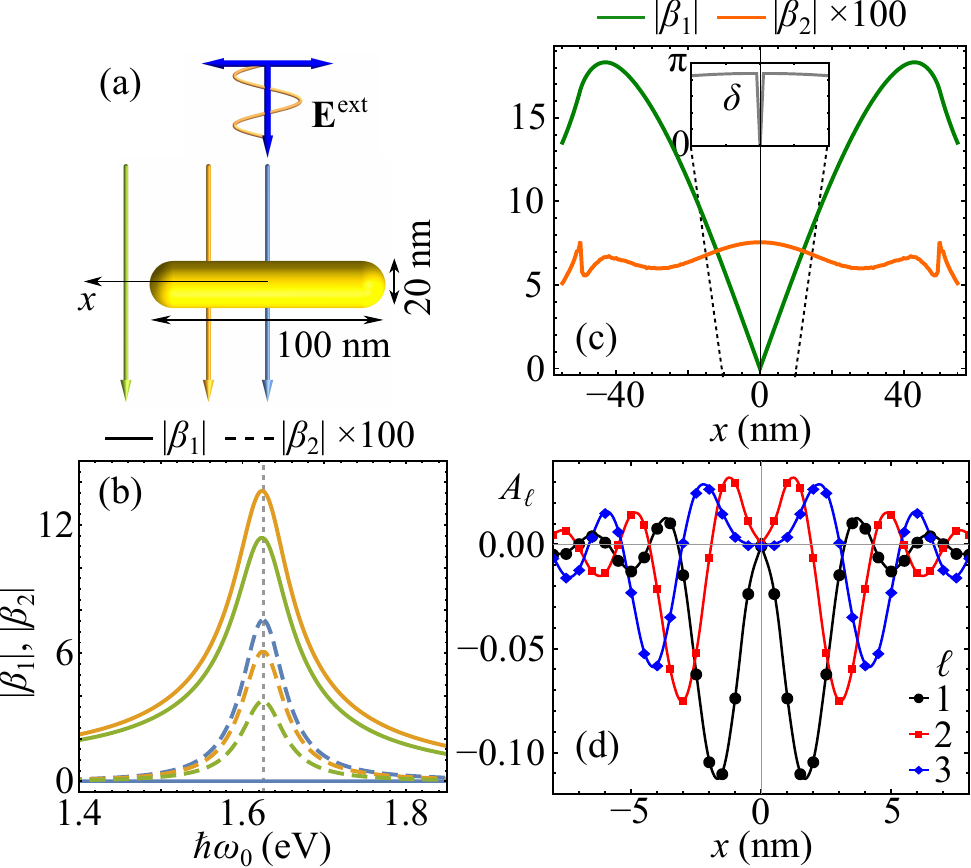}
\caption{Spatial dependence of nonlinear PINEM. (a) We consider a gold nanorod (100\,nm long, hemispherical caps) illuminated with polarization along the particle axis. (b) Absolute values of the coupling coefficients $\abs{\beta_1}$ (solid lines) and $\abs{\beta_2}$ (dashed lines) as a function of light frequency $\omega_0$ for electron beams (100\,keV) crossing the rod axis at the positions indicated by the color-coordinated downward arrows in (a), external field amplitude $\abs{\Eb^\mathrm{ext}}=5\times10^{7}$~V/m, and SH surface susceptibility $\chi_{\perp\perp\perp}=10^{-18}~\mathrm{m^2/V}$. (c) Electron-beam-position dependence of $\abs{\beta_1}$ (green) and $\abs{\beta_2}$ (orange) for $\hbar\omega_0\approx1.63$~eV on resonance with the dipolar rod plasmon. The inset shows the phase $\delta$ close to the rod center. (d) Asymmetry parameter $A_\ell$ extracted from (c) near the rod center.}
\label{Fig4}
\end{figure}
% --------------------------------------------------

\section{Effect of cancellation of linear PINEM}

By virtue of symmetry, $\beta_1$ should vanish for an electron passing through the center of a gold nanorod under the orientation and illumination conditions shown in Fig.\ \ref{Fig4}(a). By numerically calculating the $\beta_j$ coefficients as described above, we indeed observe a vanishing of $\beta_1$ for the central trajectory [Fig.\ \ref{Fig4}(b,c)], while $|\beta_2|$ takes sizeable values even for a moderate, experimentally feasible incident light amplitude $<10^8\,$V/m \cite{FES15}. When moving the electron beam away from the rod center, the coupling coefficients change significantly, but in all cases display a prominent $\approx1.63$~eV spectral feature associated with the rod dipolar plasmon [Fig.\ \ref{Fig4}(b)]. In the central trajectory, although $\beta_1=0$, the nonzero $\beta_2$ produces a symmetric spectrum with small integrated inelastic probability $\propto|\beta_2|^2$ (see above). It is therefore convenient to have a suitable nonzero value of $|\beta_1|$ to better observe nonlinear effects (see Fig.~\ref{Fig2}). Such a regime can be reached for beam positions slightly off the rod center, as shown in Fig.\ \ref{Fig4}(c) as the electron beam is scanned along the rod for illumination on resonance with the dipolar rod plasmon. Remarkably, the resulting spectral asymmetry reaches $A_1\sim-10\%$ in the first sideband [Fig.~\ref{Fig4}(d)] under the considered realistic conditions.

% \section{Conclusions} % ---------------------
\section{Concluding remarks}

In summary, our results support the use of stimulated light-electron interactions near nanostructures to locally and quantitatively probe the nonlinear response of the materials forming them. This idea can directly be implemented through careful analysis of PINEM data using existing microscope setups. Importantly, signatures of the second-harmonic response appear as contributions to the electron spectra scaling linearly with the nonlinear field amplitude, rather than its intensity. Further study is needed to explore the ability of resolving higher-order nonlinear processes. Improvements in the nonlinear detection efficiency could arise by making the electron interact with additional illuminated structures, whereby the linear coupling coefficient $\beta_1$ could be manipulated to better resolve the nonlinear contribution $\beta_2$. Combined with tomography through sample and light beam rotation and spatial sampling of the near field, more detailed information on the spatial dependence of the nonlinear response could be also obtained. Exploiting these methods, an interesting possibility is presented by quantum nonlinearites present in Jaynes-Cummings coupling \cite{JC1963} of quantum optical emitters in optical cavities. In brief, tightly focused electron beams can facilitate the determination of the optical nonlinear response for small amounts of material with unprecedented spatial resolution using currently available ultrafast electron microscopes.

\acknowledgments

We thank A. Feist, K. E. Priebe, and S.V. Yalunin for stimulating discussions. This work has been supported in part by ERC (Advanced Grant 789104-eNANO), the Spanish MINECO (MAT2017-88492-R and SEV2015-0522), the Catalan CERCA Program, Fundaci\'{o} Privada Cellex, and the Deutsche Forschungsgemeinschaft (SFB 1073, project A05). V.D.G. acknowledges support from the EU through a Marie Sk\l{}odowska-Curie grant (COFUND-DP, H2020-MSCA-COFUND-2014, GA n 665884).

\appendix

\begin{widetext}

\section{Electron wave function and spectra in nonlinear PINEM}

Following a previous theoretical formulation of PINEM interactions \cite{paper311}, we write the electron wave function as
\begin{align}
\psi(\bm{r},t)=\ee^{\ii(\mathbf{k}_0\cdot\mathbf{r}-E_0 t/\hbar)}\phi(\mathbf{r},t) \nonumber
\end{align}
for an electron of momentum and energy components tightly focused around $\hbar\mathbf{k}_0$ and $E_0$, where
\begin{align}
\phi(\mathbf{r},t)=\phi_0(\mathbf{R}_0,z,t)\,\mathrm{exp}\left[\frac{-\ii e}{\hbar c}\int_{-\infty}^t dt' A_z(\mathbf{R}_0,z+vt'-v t,t')\right],
\label{phi}
\end{align}
$\phi_0$ defines the incident wave function profile, and
\begin{align}
\mathbf{A}(\mathbf{R}_0,z,t)=\frac{-\ii c}{\omega_0}\sum\limits_{j=1}^\infty\left[\frac{1}{j}\mathbf{E}^{(j)}(\mathbf{R}_0,z,t)\ee^{-\ii j \omega_0 t}\right]+{\rm c.c.} \label{AA}
\end{align}
is the light vector potential, in which we assume continuous-wave illumination for simplicity and introduce the contribution of different harmonics $j$ due to the nonlinear response of the sample. We consider the electron to move along the $z$ direction and the electron beam to be focused at a lateral position $\Rb_0=(x_0,y_0)$ with a transversal size that is small compared with the spatial variation of any of the optical fields involved in the interaction. Upon insertion of Eq.\ (\ref{AA}) into Eq.\ (\ref{phi}), we readily find
\begin{align}
\phi(\mathbf{r},t)&=\phi_0(\mathbf{R}_0,z,t)\;\exp\left[2\ii \sum\limits_{j=1}^\infty
\abs{\beta_j(z)}\,\mathrm{sin}(\arg\{\beta_j(z)\}+j\omega_0(z/v-t))\right], \nonumber
\end{align}
where
\begin{align}
\beta_j(z)&=\frac{e}{j\hbar\omega_0}\int_{-\infty}^z dz'E_{z}^{(j)}(\mathbf{R}_0,z')\ee^{-\ii j \omega_0z'/v} \label{beta}
\end{align}
is the $j$ harmonic coupling coefficient. We now use the Jacobi-Anger relation \cite{DTL1996}
\begin{align}
\mathrm{exp}\left[(x/2)(u-1/u)\right] = \sum_{n=-\infty}^{\infty}u^n J_{n}(x)
\nonumber
\end{align}
with $u=\exp\left[\ii(\arg\{\beta_j(z)\}+j\omega_0(z/v-t))\right]$ to write
\begin{align}
\phi(\mathbf{r},t)&=\phi_0(\mathbf{R}_0,z,t)\,\prod\limits_{j=1}^{\infty}\sum_{n_j=-\infty}^{\infty} \ee^{\ii n_j(\arg\{\beta_j(z)\}+j\omega_0(z/v-t))}J_{n_j}(2\abs{\beta_j(z)}).
\nonumber
\end{align}
This expression is general and can be readily applied to include an arbitrarily large number of harmonics. In our study, we specify it to only $j=1$ (fundamental mode) and $j=2$ [second-harmonic (SH) mode], so it readily simplifies to
\begin{align}
\phi(\mathbf{r},t)&=\phi_0(\mathbf{R}_0,z,t)\,\sum\limits_{\ell=-\infty}^\infty\,f_\ell\,\ee^{\ii\ell\omega_0(z/v-t)}, \nonumber
\end{align}
where
\begin{align}
f_\ell=\ee^{\ii \ell\arg\{-\beta_1\}}\sum\limits_{n=-\infty}^\infty \ee^{-\ii n\delta} J_{\ell+2n}(2\abs{\beta_1})J_n(2\abs{\beta_2})
\label{fifinal}
\end{align}
and
\begin{align}
\delta=\arg\{\beta_2\}-2\arg\{\beta_1\}.
\label{delta}
\end{align}
We apply Eqs.\ (\ref{fifinal}) and (\ref{delta}) in the main text with the definition $\beta_j\equiv\beta_j(z=\infty)$.

\section{Asymmetry parameter}

{\it Definition.---}We define the asymmetry parameter associated with a sideband $\ell$ from the probabilities $P_\ell=|f_\ell|^2$ as
\begin{align}
A_\ell=P_\ell -P_{-\ell},
\nonumber
\end{align}
bound to the range $-1\le A_\ell\le1$ because $0\le P_\ell\le1$. However, we find them to be numerically bound to smaller ranges of sizes decreasing with increasing $\ell$ as $-0.23\lesssim A_1\lesssim0.53$, $-0.20\lesssim A_2\lesssim0.45$, $-0.16\lesssim A_3\lesssim0.38$, etc.

{\it Vanishing points.---}Interestingly, these parameters vanish simultaneously for all sidebands when the phase defined in Eq.\ (\ref{delta}) takes values $\delta=(2m+1)\pi/2$, where $m$ runs over integer numbers, as we can readily verify upon direct inspection of the expression
\begin{align}
A_{\ell}=\abs{f_\ell}^2-\abs{f_{-\ell}}^2&=\sum_{n,n'}\left[1-(-1)^{n+n'}\right]\;\alpha_{nn'}\;\cos[(n-n')\delta] \nonumber\\
&=2\sum\limits_{\substack{n,n' \\ \textrm{odd}\,\,n+n'}}\alpha_{nn'}\;\cos[(n-n')\delta], \nonumber
\end{align}
written in terms of the symmetric coefficients
\[\alpha_{nn'}=\alpha_{n'n}=J_{\ell+2n}(2\abs{\beta_1})J_{\ell+2n'}(2\abs{\beta_1})J_n(2\abs{\beta_2})J_{n'}(2\abs{\beta_2})\]
by direct application of $P_\ell=|f_\ell|^2$ with $f_\ell$ given by Eq.\ (\ref{fifinal}).

{\it Stationary points.---}The derivative of $A_\ell$ with respect to $\delta$ renders $\sin[(n-n')\delta]$ functions that obviously vanish when $\delta$ is a multiple of $\pi$.

{\it Small $\beta_2$ limit.---}For the $|\beta_2|\ll1$ values that we expect to encounter in practice (i.e., for currently existing nonlinear materials, and in particular for the gold nanoparticles considered in this work), using the series expansion $J_{n\ge0}(x)=(x/2)^n \sum_{k=0}^\infty (-x^2/4)^k/k!(n+k)!$ and the property $J_n(x)=(-1)^n J_{-n}(x)$ in Eq.\ (\ref{fifinal}), we find
\begin{align}
f_\ell=\ee^{\ii \ell\arg\{-\beta_1\}}
\bigg\{&J_\ell(2\abs{\beta_1}) \nonumber\\
&+|\beta_2| \left[ \ee^{-\ii\delta}J_{\ell+2}(2|\beta_1|) 
-\ee^{\ii\delta}J_{\ell-2}(2|\beta_1|) \right] \nonumber \\
&+|\beta_2|^2 \left[ \ee^{-2\ii\delta}J_{\ell+4}(2|\beta_1|)
+\ee^{2\ii\delta}J_{\ell-4}(2|\beta_1|)-J_\ell(2\abs{\beta_1}) \right]  \bigg\} \nonumber\\
&+O(\beta_2^3), \nonumber
\end{align}
from which we only retain up to linear terms in $\beta_2$ in the main text.

\section{Analytical linear and SH coupling coefficients for a small sphere in the quasistatic limit}
\label{SSec:Sphere}

{\it Linear field produced by an illuminated sphere.---}The general solution of the electric potential satisfying the Laplace equation in a homogeneous region of space can be written in spherical coordinates as
\begin{align}
    \phi(r,\theta,\varphi)=\sum_{l=0}^\infty\sum_{m=-l}^l\left(A_{lm} r^l+\frac{B_{lm}}{r^{l+1}}\right)Y_{lm}(\theta,\varphi), \label{figeneral}
\end{align}
where $Y_{lm}(\theta,\varphi)$ are spherical harmonics and $A_{lm}$ and $B_{lm}$ are constant expansion coefficients. We now consider a homogenous sphere of radius $a$ and permittivity $\epsilon_j$ evaluated at a frequency $j\omega_0$ (taken in general as a harmonic $j$ of the fundamental frequency $\omega_0$). This frequency is imposed by an external potential produced by sources assumed to be placed outside the particle (see below). We can separate the external potential in the sphere region as a sum over components $\phi_{lm}^{\mathrm{ext},(j)}=A_{lm}^{\mathrm{ext},(j)}r^l Y_{lm}$, for which the induced potential becomes
\begin{align}
\phi_{lm}^{\mathrm{ind},(j)}=-A_{lm}^{\mathrm{ext},(j)}\,Y_{lm}\,\alpha_l\times 
\begin{cases}
1/r^{l+1},& \quad r\geq a, \\
r^l/a^{2l+1},& \quad r<a,
\end{cases}
\label{fiindgeneral}
\end{align}
which is continuous at $r=a$ by construction and for which the coefficient
\begin{align}
    \alpha_l^{(j)}&=\frac{l(\epsilon_j-1)}{l\epsilon_j+l+1}a^{2l+1} \nonumber
\end{align}
further guarantees the continuity of $\epsilon\,\partial_r\phi$.

In particular, for external illumination with a light electric field amplitude $\Eb^{\rm ext}$, the external potential $-\rb\cdot\Eb^{\rm ext}$ can be written as in Eq.\ (\ref{figeneral}) with
\begin{subequations}
\begin{align}
&A^{\mathrm{ext},(1)}_{10}=-\sqrt{4\pi/3}\;E_z^{\rm ext}, \\
&A^{\mathrm{ext},(1)}_{11}=\sqrt{2\pi/3}\;(E_x^{\rm ext}-\ii E_y^{\rm ext}), \\
&A^{\mathrm{ext},(1)}_{1,-1}=-\sqrt{2\pi/3}\;(E_x^{\rm ext}+\ii E_y^{\rm ext})
\end{align}
\label{Alinear}
\end{subequations}
as the only nonzero coefficients.

{\it Second-harmonic field.---}We consider the second-order nonlinear response originating at the metal surface, which leads to a surface distribution $P_\sb^{(2)}\nn_\sb$ of dipoles oriented along the local surface normal $\nn_\sb$, proportional to the SH polarizability component $\chi_{\perp\perp\perp}$ and the square of the normal linear field $\nn_\sb\cdot\Eb^{(1)}(\sb)$ evaluated inside the metal right below each surface position $\sb$, as described by Eq.\ (\ref{P2}). This dipole distribution, which must be in turn regarded as a field source at frequency $2\omega_0$ placed immediately outside the metal surface, is fully equivalent to the combination of two distributions of opposite charges sitting at spherical surfaces of radii $r=a$ (surface charge density $-P_\sb^{(2)}/h$) and $r=a+h$ (surface charge density $[a^2/(a+h)^2]\,P_\sb^{(2)}/h$), separated by a vanishingly small distance $h\rightarrow0$. We note that a factor $a^2/(a+h)^2$ is introduced in the outer surface charge distribution in order to preserve charge neutrality. Now, expanding
\begin{align}
P_\sb^{(2)}=\sum_{lm} P_{lm}^{(2)}Y_{lm}
\label{Plm}
\end{align}
in terms of spherical harmonics, we can write the two spherical surface charge distributions as $\sigma^\pm(\sb)=\sum_{lm} \sigma_{lm}^\pm Y_{lm}$ with coefficients
\begin{align}
\sigma_{lm}^+&=\frac{P_{lm}^{(2)}}{h}\frac{a^2}{(a+h)^2}, \nonumber\\
\sigma_{lm}^-&=-\frac{P_{lm}^{(2)}}{h}.
\nonumber
\end{align}
at $r=a$ (for $\sigma^-$) and $r=a+h$ (for $\sigma^+$).

At this point, it is useful to recall that a spherical surface charge distribution of coefficients $\sigma_{lm}$ placed at $r=b$ in vacuum produces a potential \cite{J99,paper149}
\begin{align}
\phi^{\rm ext}=\sum_{lm}\frac{4\pi\sigma_{lm}}{2l+1}\,\frac{r_<^l}{r_>^{l-1}}\,Y_{lm},
\nonumber
\end{align}
where $r_<={\rm min}\{r,b\}$ and $r_>={\rm max}\{r,b\}$. In the presence of a sphere of radius $a<b$ and permittivity $\epsilon$, an induced potential $\phi^{\rm ind}$ is generated by reflection of the $r<b$ components at the sphere surface. Using Eqs.\ (\ref{figeneral}) and (\ref{fiindgeneral}), we find
\begin{align}
\phi^{\rm ind}=\sum_{lm}\frac{4\pi\sigma_{lm}}{2l+1}\,
\frac{\alpha_l}{b^{l-1}r^{l+1}}\,Y_{lm},&\quad r>a,
\nonumber
\end{align}
therefore generating a total potential
\begin{align}
\phi=\sum_{lm}\frac{4\pi\sigma_{lm}}{2l+1}\left(b^{l+2}-\frac{\alpha_l}{b^{l-1}}\right)\,\frac{1}{r^{l+1}}\,Y_{lm}
\label{onesurface}
\end{align}
in the $r>b$ region.

Summing Eq.\ (\ref{onesurface}) for the two closely spaced spherical charge distributions discussed above and taking the $h\rightarrow0$ limit, we obtain the total potential generated outside the particle by the SH dipoles as
\begin{align}
\phi^{(2)}&=\lim_{h\rightarrow0}\sum_{lm}\frac{4\pi P_{lm}^{(2)}}{2l+1}\frac{1}{h}\left[\frac{a^2}{(a+h)^2}\left((a+h)^{l+2}-\frac{\alpha_l^{(2)}}{(a+h)^{l-1}}\right)-\left(a^{l+2}-\frac{\alpha_l^{(2)}}{a^{l-1}}\right)\right] \,\frac{1}{r^{l+1}}\,Y_{lm} \nonumber\\
&=\sum_{lm}\frac{4\pi P_{lm}^{(2)}}{2l+1}\left(la^{l+1}+(l+1)\frac{\alpha_l^{(2)}}{a^l}\right) 
\,\frac{1}{r^{l+1}}\,Y_{lm} \nonumber\\
&=\sum_{lm}4\pi l\, P_{lm}^{(2)}\;\frac{\epsilon_2}{l\epsilon_2+l+1}\,\left(\frac{a}{r}\right)^{l+1}\,Y_{lm}, \label{fi2final}
\end{align}
which includes the response of the sphere through its permittivity $\epsilon_2$ evaluated at the SH frequency $2\omega_0$.

For illumination with a light plane wave, noticing that the sphere normal is just a radial vector $\nn_\sb=\hat{\sb}$ and that $\nn_\sb\cdot\Eb^{\rm ext}$ is then $-1/a$ times the potential at the sphere surface, we find the expansion coefficients of Eq.\ (\ref{Plm}) upon projection on spherical harmonics to reduce to
\begin{align}
P_{lm}^{(2)}&= \frac{9\,\chi_{\perp\perp\perp}}{(\epsilon_1+2)^2} \frac{1}{a^2} \int d\Omega \; Y_{lm}^*(\Omega) \left(\sb\cdot\Eb^{\rm ext}\right)^2 \nonumber\\
&=\frac{9\,\chi_{\perp\perp\perp}}{(\epsilon_1+2)^2} \sum_{m'} A^{{\rm ext},(1)}_{1m'}A^{{\rm ext},(1)}_{1m-m'} \int d\Omega \; Y_{lm}^*(\Omega)Y_{1m'}(\Omega)Y_{1m-m'}(\Omega) \nonumber\\
&=\frac{9\,\chi_{\perp\perp\perp}}{(\epsilon_1+2)^2} \bigg\{\delta_{lm,00}\frac{\sqrt{4\pi}}{3}\Eb^{\rm ext}\cdot\Eb^{\rm ext}+
\delta_{l2}\frac{1}{\sqrt{20\pi}}
\sum_{m'} g_{m',m-m'} A^{{\rm ext},(1)}_{1m'}A^{{\rm ext},(1)}_{1m-m'}
\bigg\}, \nonumber
\end{align}
where the overall factor $3/(\epsilon_1+2)$ entering this expression twice corresponds to the ratio of the sub-surface field $\Eb^{(1)}$ to the external field $\Eb^{\rm ext}$. Also, the coefficients $g_{m_1m_2}=g_{m_2m_1}=g_{-m_1,-m_2}$ take the values $g_{11}=\sqrt{6}$, $g_{01}=\sqrt{3}$, $g_{-11}=1$, and $g_{00}=2$, as obtained by explicitly evaluating the integral $\int d\Omega \; Y_{lm}^*(\Omega)Y_{l'm'}(\Omega)Y_{l''m''}(\Omega)=\sqrt{(2l'+1)(2l''+1)/4\pi(2l+1)}C_{l'l''l}^{000}C_{l'l''l}^{m'm''m''}$ in terms of the Clebsch-Gordan coefficients $C$ \cite{M1966}. In particular, for $\Eb^{\rm ext}\parallel\zz$, this equation reduces to
\begin{align}
P_{lm}^{(2)}=\frac{3\sqrt{4\pi}}{(\epsilon_1+2)^2}\; \chi_{\perp\perp\perp}\,(E^{\rm ext})^2
\left(\delta_{lm,00}+
\delta_{lm,20}\frac{1}{\sqrt{5}}
\right), \nonumber
\end{align}
which upon substitution into Eq.\ (\ref{fi2final}) permits us to write
\begin{align}
\phi^{(2)}=24\pi a^3\,\frac{\epsilon_2}{(\epsilon_1+2)^2(2\epsilon_2+3)}\;\chi_{\perp\perp\perp}\,(E^{\rm ext})^2 \left(\frac{3z^2-r^2}{r^5}\right) \label{phi2}
\end{align}
for the total SH potential.

{\it Coupling coefficients.---}In the quasistatic limit, it is convenient to express the field of Eq.\ (\ref{beta}) as the gradient of the electric potential and then integrate by parts to write
\begin{align}
\beta_j=\frac{\ii e}{\hbar v}\int_{-\infty}^\infty dx\;\phi^{(j)}(x,0,b)\ee^{-\ii j \omega_0x/v}, \label{betafi}
\end{align}
where we now consider the electron velocity to be along $x$ instead of $z$ and the electron beam to pass at a distance $b$ from the sphere center (i.e., the conditions of Fig.\ \ref{Fig3}). Then, taking the external light field along $z$ again and using the fact that only the induced field couples to the electron, we obtain the linear PINEM coefficient by inserting Eqs.\ (\ref{Alinear}) into Eq.\ (\ref{fiindgeneral}), and this in turn into Eq.\ (\ref{betafi}), to yield
\begin{align}
\beta_1&=\frac{\ii e a^3}{\hbar v}\frac{\epsilon_1-1}{\epsilon_1+2}\,E^{\rm ext}\int_{-\infty}^\infty dx\;\ee^{-\ii \omega_0x/v}
\frac{z}{(x^2+z^2)^{3/2}}
 \nonumber\\
&=\frac{2 \mathrm{i} e \omega_0 a^3}{\hbar v^2}\frac{\epsilon_1-1}{\epsilon_1+2}\,E^{\rm ext}\,K_1\left(\frac{\omega_0 b}{v}\right), \label{Eq:beta1_perpendicular}
\end{align}
where the second line is derived from the first one by identifying the integral with the $\partial_z$ derivative of $\int_{-\infty}^\infty dx\;\ee^{-\ii \omega_0x/v}/\sqrt{x^2+z^2}=2K_0(\omega_0 z/v)$ \cite{GR1980}.

Likewise, the SH coupling coefficient is obtained by inserting Eq.\ (\ref{phi2}) into Eq.\ (\ref{betafi}) and identifying the integral again with successive derivatives of the $K_0$ modified Bessel function. We find
\begin{align}
\beta_2&=\frac{96\pi\mathrm{i}e \omega_0 a^3}{\hbar v^2b} \frac{\epsilon_2}{(\epsilon_1+2)^2(2\epsilon_2+3)}\;\chi_{\perp\perp\perp}\,(E^{\rm ext})^2
\left[\frac{2\omega_0b}{v}K_2\left(\frac{2\omega_0 b}{v}\right)-K_1\left(\frac{2\omega_0 b}{v}\right)\right], \label{Eq:beta2_perpendicular}
\end{align}
which scales linearly with the sphere volume. Under grazing incidence ($b=a$), aside from the dependence on the dielectric response ($\epsilon_1$, $\epsilon_2$, and $\chi_{\perp\perp\perp}$), this expression depends on $a$, $\omega_0$, and $v$ as $(1/\omega_0)f(\omega_0a/v)$, where the function $f(\theta)=2\theta^3K_2(2\theta)-\theta^2 K_1(2\theta)$ takes a maximum value $\sim0.4$ for $\omega_0a/v\sim1$.

We use Eqs.\ (\ref{Eq:beta1_perpendicular}) and (\ref{Eq:beta2_perpendicular}) to produce the linear and nonlinear calculations shown in Fig.\ \ref{Fig3}.

\section{Numerical calculation of coupling coefficients including retardation}

The retarded calculations of Fig.\ 4 are obtained based on two numerical simulations of the linear Maxwell equations at frequencies $\omega_0$ and $2\omega_0$ for sources corresponding to the external light and the SH surface dipole distribution, which produce the fields $\Eb^{(1)}$ and $\Eb^{(2)}$, respectively. We use a finite-elements methods in the frequency domain implemented in COMSOL to obtain those fields, the integration of which yields the coupling coefficients $\beta_j$ according to Eq.~\eqref{beta}.

We find it conveninent to use an alternative, faster approach to calculate $\beta_2$, based on the transformations
\begin{align}
    \beta_2&=\frac{e}{2\hbar\omega_0}\int dz\,E_z^{(2)}(\Rb_0,z)\ee^{-\mathrm{i}2\omega_0 z/v}\nonumber\\
    &=\frac{e}{2\hbar\omega_0}\int dz\int_Sd^2\mathbf{s}\;\hat{\bf z}\cdot\mathcal{G}(\Rb_0,z,\mathbf{s},2\omega_0)\cdot \mathbf{P}_\mathbf{s}^{(2)}\ee^{-\mathrm{i}2\omega_0 z/v}\nonumber\\
    &=\frac{e}{2\hbar\omega_0}\int_S d^2\mathbf{s}\, \mathbf{P}_\mathbf{s}^{(2)}\cdot\int dz\;\mathcal{G}(\mathbf{s},\Rb_0,z,2\omega_0)\cdot\hat{\bf z}\; \ee^{-\mathrm{i}2\omega_0 z/v}\nonumber\\
     &=\frac{\mathrm{i}}{2\hbar\omega_0}\int_S d^2\mathbf{s}\; \mathbf{P}_\mathbf{s}^{(2)}\cdot\mathbf{E}^\mathrm{current}(\mathbf{s},2\omega_0), \nonumber
\end{align}
where the second line expresses the SH field produced by the surface dipole distribution $\mathbf{P}_\mathbf{s}^{(2)}$ as an integral over the particle surface $S$ using the electromagnetic Green tensor $\mathcal{G}$ evaluated at frequency $2\omega_0$ (see Ref.\ \cite{paper149} for a definition that uses the same notation and Gaussian units as in the present work), we then use the reciprocity theorem to write the third line, and in the last step we identify the $z$ integral with the total field $\mathbf{E}^\mathrm{current}(\mathbf{s},2\omega_0)$ generated at the surface position $\mathbf{s}$ by an electric current corresponding to the classical $2\omega_0$ component of the electron beam oscillating as $\ee^{-\mathrm{i}2\omega_0 z/v}$ along the electron trajectory (incidentally, this field is evaluated inside the metal right beneath the surface). We thus need to perform a single electromagnetic simulation [i.e., we obtain $\mathbf{E}^\mathrm{current}(\mathbf{s},2\omega_0)$ for a line dipole source before carrying out the $\sb$ integral] instead of one simulation for the nonlinear dipole $\mathbf{P}_\mathbf{s}^{(2)}\Delta\sb$ at each surface element $\Delta\sb$.

% FIGURE S1 -----------------------------------------
\begin{figure}[h!]
    \centering
    \includegraphics[width=0.8\textwidth]{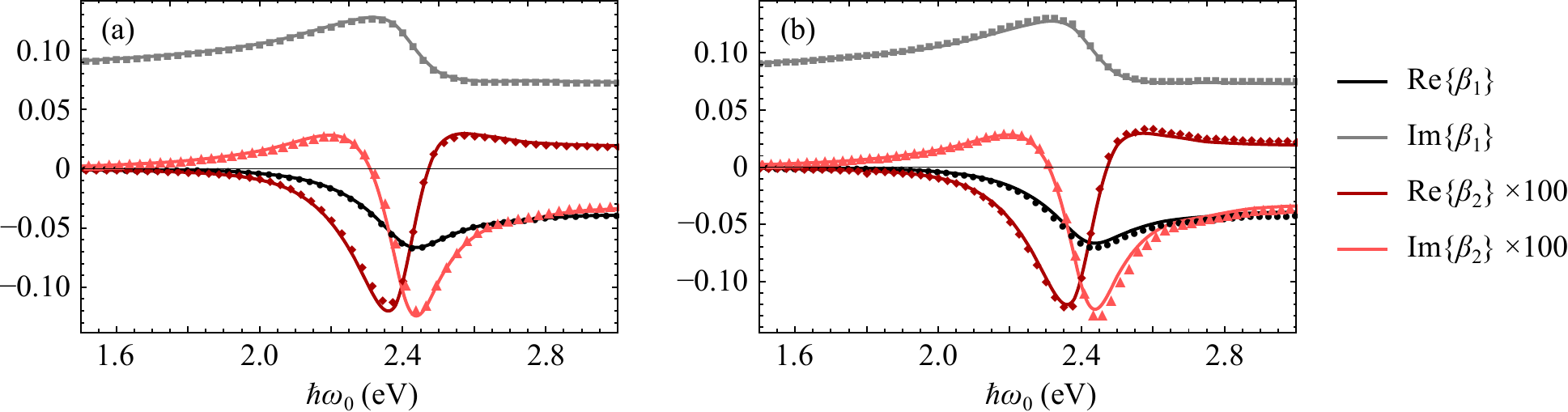}
    \caption{Comparison of analytical and numerical solutions for the sphere of Fig.~3. We show the real (dark) and imaginary (light) parts of the coupling coefficients $\beta_1$ (black and gray) and $\beta_2$ (red) as obtained from quasistatic numerical simulations [symbols in (a)], retarded numerical simulations [symbols in (b)], and analytical theory [solid curves in both (a) and (b), obtained from Eqs.~\eqref{Eq:beta1_perpendicular} and \eqref{Eq:beta2_perpendicular}]. All calculations are carried out under the same conditions as in Fig.~3.}
    \label{FigS1}
\end{figure}

We compare our analytical expressions for the coupling coefficients [Eqs.~\eqref{Eq:beta1_perpendicular} and \eqref{Eq:beta2_perpendicular}] with numerical solutions in Fig.\ \ref{FigS1} for a sphere with the same parameters as in Fig.\ \ref{Fig3}. Analytical and numerical results in the quastistatic limit are in nearly perfect agreement [Fig.\ \ref{FigS1}(a)]. When including retardation, the numerical results are still in good agreement, only showing minor discrepancies with respect to the quasistatic analytical results [Fig.\ \ref{FigS1}(b)]. This confirms that retardation does not play a significant role for the relatively small particle under consideration.

\end{widetext}

%\bibliographystyle{apsrev}
%\bibliography{../../../bibtex/refs}

\begin{thebibliography}{66}
\expandafter\ifx\csname natexlab\endcsname\relax\def\natexlab#1{#1}\fi
\expandafter\ifx\csname bibnamefont\endcsname\relax
  \def\bibnamefont#1{#1}\fi
\expandafter\ifx\csname bibfnamefont\endcsname\relax
  \def\bibfnamefont#1{#1}\fi
\expandafter\ifx\csname citenamefont\endcsname\relax
  \def\citenamefont#1{#1}\fi
\expandafter\ifx\csname url\endcsname\relax
  \def\url#1{\texttt{#1}}\fi
\expandafter\ifx\csname urlprefix\endcsname\relax\def\urlprefix{URL }\fi
\providecommand{\bibinfo}[2]{#2}
\providecommand{\eprint}[2][]{\url{#2}}

\bibitem[{\citenamefont{Egerton}(1996)}]{E96}
\bibinfo{author}{\bibfnamefont{R.~F.} \bibnamefont{Egerton}},
  \emph{\bibinfo{title}{Electron Energy-Loss Spectroscopy in the Electron
  Microscope}} (\bibinfo{publisher}{Plenum Press}, \bibinfo{address}{New York},
  \bibinfo{year}{1996}).

\bibitem[{\citenamefont{Barwick et~al.}(2009)\citenamefont{Barwick, Flannigan,
  and Zewail}}]{BFZ09}
\bibinfo{author}{\bibfnamefont{B.}~\bibnamefont{Barwick}},
  \bibinfo{author}{\bibfnamefont{D.~J.} \bibnamefont{Flannigan}},
  \bibnamefont{and} \bibinfo{author}{\bibfnamefont{A.~H.}
  \bibnamefont{Zewail}}, \bibinfo{journal}{Nature}
  \textbf{\bibinfo{volume}{462}}, \bibinfo{pages}{902} (\bibinfo{year}{2009}).

\bibitem[{\citenamefont{{Garc\'{\i}a de Abajo}
  et~al.}(2010)\citenamefont{{Garc\'{\i}a de Abajo}, {Asenjo Garcia}, and
  Kociak}}]{paper151}
\bibinfo{author}{\bibfnamefont{F.~J.} \bibnamefont{{Garc\'{\i}a de Abajo}}},
  \bibinfo{author}{\bibfnamefont{A.}~\bibnamefont{{Asenjo Garcia}}},
  \bibnamefont{and} \bibinfo{author}{\bibfnamefont{M.}~\bibnamefont{Kociak}},
  \bibinfo{journal}{Nano\ Lett.} \textbf{\bibinfo{volume}{10}},
  \bibinfo{pages}{1859} (\bibinfo{year}{2010}).

\bibitem[{\citenamefont{Park et~al.}(2010)\citenamefont{Park, Lin, and
  Zewail}}]{PLZ10}
\bibinfo{author}{\bibfnamefont{S.~T.} \bibnamefont{Park}},
  \bibinfo{author}{\bibfnamefont{M.}~\bibnamefont{Lin}}, \bibnamefont{and}
  \bibinfo{author}{\bibfnamefont{A.~H.} \bibnamefont{Zewail}},
  \bibinfo{journal}{New\ J.\ Phys.} \textbf{\bibinfo{volume}{12}},
  \bibinfo{pages}{123028} (\bibinfo{year}{2010}).

\bibitem[{\citenamefont{Krivanek et~al.}(2014)\citenamefont{Krivanek, Lovejoy,
  Dellby, Aoki, Carpenter, Rez, Soignard, Zhu, Batson, Lagos et~al.}}]{KLD14}
\bibinfo{author}{\bibfnamefont{O.~L.} \bibnamefont{Krivanek}},
  \bibinfo{author}{\bibfnamefont{T.~C.} \bibnamefont{Lovejoy}},
  \bibinfo{author}{\bibfnamefont{N.}~\bibnamefont{Dellby}},
  \bibinfo{author}{\bibfnamefont{T.}~\bibnamefont{Aoki}},
  \bibinfo{author}{\bibfnamefont{R.~W.} \bibnamefont{Carpenter}},
  \bibinfo{author}{\bibfnamefont{P.}~\bibnamefont{Rez}},
  \bibinfo{author}{\bibfnamefont{E.}~\bibnamefont{Soignard}},
  \bibinfo{author}{\bibfnamefont{J.}~\bibnamefont{Zhu}},
  \bibinfo{author}{\bibfnamefont{P.~E.} \bibnamefont{Batson}},
  \bibinfo{author}{\bibfnamefont{M.~J.} \bibnamefont{Lagos}},
  \bibnamefont{et~al.}, \bibinfo{journal}{Nature}
  \textbf{\bibinfo{volume}{514}}, \bibinfo{pages}{209} (\bibinfo{year}{2014}).

\bibitem[{\citenamefont{Kirchner et~al.}(2014)\citenamefont{Kirchner, Gliserin,
  Krausz, and Baum}}]{KGK14}
\bibinfo{author}{\bibfnamefont{F.~O.} \bibnamefont{Kirchner}},
  \bibinfo{author}{\bibfnamefont{A.}~\bibnamefont{Gliserin}},
  \bibinfo{author}{\bibfnamefont{F.}~\bibnamefont{Krausz}}, \bibnamefont{and}
  \bibinfo{author}{\bibfnamefont{P.}~\bibnamefont{Baum}},
  \bibinfo{journal}{Nat.\ Photon.} \textbf{\bibinfo{volume}{8}},
  \bibinfo{pages}{52} (\bibinfo{year}{2014}).

\bibitem[{\citenamefont{Feist et~al.}(2015)\citenamefont{Feist, Echternkamp,
  Schauss, Yalunin, Sch\"afer, and Ropers}}]{FES15}
\bibinfo{author}{\bibfnamefont{A.}~\bibnamefont{Feist}},
  \bibinfo{author}{\bibfnamefont{K.~E.} \bibnamefont{Echternkamp}},
  \bibinfo{author}{\bibfnamefont{J.}~\bibnamefont{Schauss}},
  \bibinfo{author}{\bibfnamefont{S.~V.} \bibnamefont{Yalunin}},
  \bibinfo{author}{\bibfnamefont{S.}~\bibnamefont{Sch\"afer}},
  \bibnamefont{and} \bibinfo{author}{\bibfnamefont{C.}~\bibnamefont{Ropers}},
  \bibinfo{journal}{Nature} \textbf{\bibinfo{volume}{521}},
  \bibinfo{pages}{200} (\bibinfo{year}{2015}).

\bibitem[{\citenamefont{Piazza et~al.}(2015)\citenamefont{Piazza, Lummen,
  {Qui\~{n}onez}, Murooka, Reed, Barwick, and Carbone}}]{PLQ15}
\bibinfo{author}{\bibfnamefont{L.}~\bibnamefont{Piazza}},
  \bibinfo{author}{\bibfnamefont{T.~T.~A.} \bibnamefont{Lummen}},
  \bibinfo{author}{\bibfnamefont{E.}~\bibnamefont{{Qui\~{n}onez}}},
  \bibinfo{author}{\bibfnamefont{Y.}~\bibnamefont{Murooka}},
  \bibinfo{author}{\bibfnamefont{B.}~\bibnamefont{Reed}},
  \bibinfo{author}{\bibfnamefont{B.}~\bibnamefont{Barwick}}, \bibnamefont{and}
  \bibinfo{author}{\bibfnamefont{F.}~\bibnamefont{Carbone}},
  \bibinfo{journal}{Nat.\ Commun.} \textbf{\bibinfo{volume}{6}},
  \bibinfo{pages}{6407} (\bibinfo{year}{2015}).

\bibitem[{\citenamefont{Echternkamp et~al.}(2016)\citenamefont{Echternkamp,
  Feist, Sch\"{a}fer, and Ropers}}]{EFS16}
\bibinfo{author}{\bibfnamefont{K.~E.} \bibnamefont{Echternkamp}},
  \bibinfo{author}{\bibfnamefont{A.}~\bibnamefont{Feist}},
  \bibinfo{author}{\bibfnamefont{S.}~\bibnamefont{Sch\"{a}fer}},
  \bibnamefont{and} \bibinfo{author}{\bibfnamefont{C.}~\bibnamefont{Ropers}},
  \bibinfo{journal}{Nat.\ Phys.} \textbf{\bibinfo{volume}{12}},
  \bibinfo{pages}{1000} (\bibinfo{year}{2016}).

\bibitem[{\citenamefont{Ryabov and Baum}(2016)}]{RB16}
\bibinfo{author}{\bibfnamefont{A.}~\bibnamefont{Ryabov}} \bibnamefont{and}
  \bibinfo{author}{\bibfnamefont{P.}~\bibnamefont{Baum}},
  \bibinfo{journal}{Science} \textbf{\bibinfo{volume}{353}},
  \bibinfo{pages}{374} (\bibinfo{year}{2016}).

\bibitem[{\citenamefont{Vanacore et~al.}(2016)\citenamefont{Vanacore,
  Fitzpatrick, and Zewail}}]{VFZ16}
\bibinfo{author}{\bibfnamefont{G.~M.} \bibnamefont{Vanacore}},
  \bibinfo{author}{\bibfnamefont{A.~W.~P.} \bibnamefont{Fitzpatrick}},
  \bibnamefont{and} \bibinfo{author}{\bibfnamefont{A.~H.}
  \bibnamefont{Zewail}}, \bibinfo{journal}{Nano\ Today}
  \textbf{\bibinfo{volume}{11}}, \bibinfo{pages}{228} (\bibinfo{year}{2016}).

\bibitem[{\citenamefont{Koz\'ak et~al.}(2017)\citenamefont{Koz\'ak, McNeur,
  Leedle, Deng, Sch\"onenberger, Ruehl, Hartl, Harris, Byer, and
  Hommelhoff}}]{KML17}
\bibinfo{author}{\bibfnamefont{M.}~\bibnamefont{Koz\'ak}},
  \bibinfo{author}{\bibfnamefont{J.}~\bibnamefont{McNeur}},
  \bibinfo{author}{\bibfnamefont{K.~J.} \bibnamefont{Leedle}},
  \bibinfo{author}{\bibfnamefont{H.}~\bibnamefont{Deng}},
  \bibinfo{author}{\bibfnamefont{N.}~\bibnamefont{Sch\"onenberger}},
  \bibinfo{author}{\bibfnamefont{A.}~\bibnamefont{Ruehl}},
  \bibinfo{author}{\bibfnamefont{I.}~\bibnamefont{Hartl}},
  \bibinfo{author}{\bibfnamefont{J.~S.} \bibnamefont{Harris}},
  \bibinfo{author}{\bibfnamefont{R.~L.} \bibnamefont{Byer}}, \bibnamefont{and}
  \bibinfo{author}{\bibfnamefont{P.}~\bibnamefont{Hommelhoff}},
  \bibinfo{journal}{Nat.\ Commun.} \textbf{\bibinfo{volume}{8}},
  \bibinfo{pages}{14342} (\bibinfo{year}{2017}).

\bibitem[{\citenamefont{Feist et~al.}(2017)\citenamefont{Feist, Bach,
  N.~Rubiano~{da Silva}, M\"{a}ller, Priebe, Domr\"{a}se, Gatzmann, Rost,
  Schauss, Strauch et~al.}}]{FBR17}
\bibinfo{author}{\bibfnamefont{A.}~\bibnamefont{Feist}},
  \bibinfo{author}{\bibfnamefont{N.}~\bibnamefont{Bach}},
  \bibinfo{author}{\bibfnamefont{T.~D.} \bibnamefont{N.~Rubiano~{da Silva}}},
  \bibinfo{author}{\bibfnamefont{M.}~\bibnamefont{M\"{a}ller}},
  \bibinfo{author}{\bibfnamefont{K.~E.} \bibnamefont{Priebe}},
  \bibinfo{author}{\bibfnamefont{T.}~\bibnamefont{Domr\"{a}se}},
  \bibinfo{author}{\bibfnamefont{J.~G.} \bibnamefont{Gatzmann}},
  \bibinfo{author}{\bibfnamefont{S.}~\bibnamefont{Rost}},
  \bibinfo{author}{\bibfnamefont{J.}~\bibnamefont{Schauss}},
  \bibinfo{author}{\bibfnamefont{S.}~\bibnamefont{Strauch}},
  \bibnamefont{et~al.}, \bibinfo{journal}{Ultramicroscopy}
  \textbf{\bibinfo{volume}{176}}, \bibinfo{pages}{63} (\bibinfo{year}{2017}).

\bibitem[{\citenamefont{Lagos et~al.}(2017)\citenamefont{Lagos, Tr\"ugler,
  Hohenester, and Batson}}]{LTH17}
\bibinfo{author}{\bibfnamefont{M.~J.} \bibnamefont{Lagos}},
  \bibinfo{author}{\bibfnamefont{A.}~\bibnamefont{Tr\"ugler}},
  \bibinfo{author}{\bibfnamefont{U.}~\bibnamefont{Hohenester}},
  \bibnamefont{and} \bibinfo{author}{\bibfnamefont{P.~E.}
  \bibnamefont{Batson}}, \bibinfo{journal}{Nature}
  \textbf{\bibinfo{volume}{543}}, \bibinfo{pages}{529} (\bibinfo{year}{2017}).

\bibitem[{\citenamefont{Priebe et~al.}(2017)\citenamefont{Priebe, Rathje,
  Yalunin, Hohage, Feist, Sch\"{a}fer, and Ropers}}]{PRY17}
\bibinfo{author}{\bibfnamefont{K.~E.} \bibnamefont{Priebe}},
  \bibinfo{author}{\bibfnamefont{C.}~\bibnamefont{Rathje}},
  \bibinfo{author}{\bibfnamefont{S.~V.} \bibnamefont{Yalunin}},
  \bibinfo{author}{\bibfnamefont{T.}~\bibnamefont{Hohage}},
  \bibinfo{author}{\bibfnamefont{A.}~\bibnamefont{Feist}},
  \bibinfo{author}{\bibfnamefont{S.}~\bibnamefont{Sch\"{a}fer}},
  \bibnamefont{and} \bibinfo{author}{\bibfnamefont{C.}~\bibnamefont{Ropers}},
  \bibinfo{journal}{Nat.\ Photon.} \textbf{\bibinfo{volume}{11}},
  \bibinfo{pages}{793} (\bibinfo{year}{2017}).

\bibitem[{\citenamefont{Pomarico et~al.}(2018)\citenamefont{Pomarico, Madan,
  Berruto, Vanacore, Wang, Kaminer, {Garc\'{\i}a de Abajo}, and
  Carbone}}]{paper306}
\bibinfo{author}{\bibfnamefont{E.}~\bibnamefont{Pomarico}},
  \bibinfo{author}{\bibfnamefont{I.}~\bibnamefont{Madan}},
  \bibinfo{author}{\bibfnamefont{G.}~\bibnamefont{Berruto}},
  \bibinfo{author}{\bibfnamefont{G.~M.} \bibnamefont{Vanacore}},
  \bibinfo{author}{\bibfnamefont{K.}~\bibnamefont{Wang}},
  \bibinfo{author}{\bibfnamefont{I.}~\bibnamefont{Kaminer}},
  \bibinfo{author}{\bibfnamefont{F.~J.} \bibnamefont{{Garc\'{\i}a de Abajo}}},
  \bibnamefont{and} \bibinfo{author}{\bibfnamefont{F.}~\bibnamefont{Carbone}},
  \bibinfo{journal}{ACS\ Photon.} \textbf{\bibinfo{volume}{5}},
  \bibinfo{pages}{759} (\bibinfo{year}{2018}).

\bibitem[{\citenamefont{Vanacore et~al.}(2018)\citenamefont{Vanacore, Madan,
  Berruto, Wang, Pomarico, Lamb, McGrouther, Kaminer, Barwick, {Garc\'{\i}a de
  Abajo} et~al.}}]{paper311}
\bibinfo{author}{\bibfnamefont{G.~M.} \bibnamefont{Vanacore}},
  \bibinfo{author}{\bibfnamefont{I.}~\bibnamefont{Madan}},
  \bibinfo{author}{\bibfnamefont{G.}~\bibnamefont{Berruto}},
  \bibinfo{author}{\bibfnamefont{K.}~\bibnamefont{Wang}},
  \bibinfo{author}{\bibfnamefont{E.}~\bibnamefont{Pomarico}},
  \bibinfo{author}{\bibfnamefont{R.~J.} \bibnamefont{Lamb}},
  \bibinfo{author}{\bibfnamefont{D.}~\bibnamefont{McGrouther}},
  \bibinfo{author}{\bibfnamefont{I.}~\bibnamefont{Kaminer}},
  \bibinfo{author}{\bibfnamefont{B.}~\bibnamefont{Barwick}},
  \bibinfo{author}{\bibfnamefont{F.~J.} \bibnamefont{{Garc\'{\i}a de Abajo}}},
  \bibnamefont{et~al.}, \bibinfo{journal}{Nat.\ Commun.}
  \textbf{\bibinfo{volume}{9}}, \bibinfo{pages}{2694} (\bibinfo{year}{2018}).

\bibitem[{\citenamefont{Vanacore et~al.}(2019)\citenamefont{Vanacore, Berruto,
  Madan, Pomarico, Biagioni, Lamb, McGrouther, Reinhardt, Kaminer, Barwick
  et~al.}}]{paper332}
\bibinfo{author}{\bibfnamefont{G.~M.} \bibnamefont{Vanacore}},
  \bibinfo{author}{\bibfnamefont{G.}~\bibnamefont{Berruto}},
  \bibinfo{author}{\bibfnamefont{I.}~\bibnamefont{Madan}},
  \bibinfo{author}{\bibfnamefont{E.}~\bibnamefont{Pomarico}},
  \bibinfo{author}{\bibfnamefont{P.}~\bibnamefont{Biagioni}},
  \bibinfo{author}{\bibfnamefont{R.~J.} \bibnamefont{Lamb}},
  \bibinfo{author}{\bibfnamefont{D.}~\bibnamefont{McGrouther}},
  \bibinfo{author}{\bibfnamefont{O.}~\bibnamefont{Reinhardt}},
  \bibinfo{author}{\bibfnamefont{I.}~\bibnamefont{Kaminer}},
  \bibinfo{author}{\bibfnamefont{B.}~\bibnamefont{Barwick}},
  \bibnamefont{et~al.}, \bibinfo{journal}{Nat.\ Mater.}
  \textbf{\bibinfo{volume}{18}}, \bibinfo{pages}{573} (\bibinfo{year}{2019}).

\bibitem[{\citenamefont{Wang et~al.}(2019)\citenamefont{Wang, Dahan, Shentcis,
  Kauffmann, Tsesses, , and Kaminer}}]{WDS19}
\bibinfo{author}{\bibfnamefont{K.}~\bibnamefont{Wang}},
  \bibinfo{author}{\bibfnamefont{R.}~\bibnamefont{Dahan}},
  \bibinfo{author}{\bibfnamefont{M.}~\bibnamefont{Shentcis}},
  \bibinfo{author}{\bibfnamefont{Y.}~\bibnamefont{Kauffmann}},
  \bibinfo{author}{\bibfnamefont{S.}~\bibnamefont{Tsesses}}, ,
  \bibnamefont{and} \bibinfo{author}{\bibfnamefont{I.}~\bibnamefont{Kaminer}},
  p. \bibinfo{pages}{arXiv:1908.06206} (\bibinfo{year}{2019}).

\bibitem[{\citenamefont{Kfir et~al.}(2019)\citenamefont{Kfir,
  Louren\c{c}o-Martins, Storeck, Sivis, Harvey, Kippenberg, Feist, and
  Ropers}}]{KLS19}
\bibinfo{author}{\bibfnamefont{O.}~\bibnamefont{Kfir}},
  \bibinfo{author}{\bibfnamefont{H.}~\bibnamefont{Louren\c{c}o-Martins}},
  \bibinfo{author}{\bibfnamefont{G.}~\bibnamefont{Storeck}},
  \bibinfo{author}{\bibfnamefont{M.}~\bibnamefont{Sivis}},
  \bibinfo{author}{\bibfnamefont{T.~R.} \bibnamefont{Harvey}},
  \bibinfo{author}{\bibfnamefont{T.~J.} \bibnamefont{Kippenberg}},
  \bibinfo{author}{\bibfnamefont{A.}~\bibnamefont{Feist}}, \bibnamefont{and}
  \bibinfo{author}{\bibfnamefont{C.}~\bibnamefont{Ropers}}, p.
  \bibinfo{pages}{arXiv:1910.09540} (\bibinfo{year}{2019}).

\bibitem[{\citenamefont{Dahan et~al.}(2019)\citenamefont{Dahan, Nehemia,
  Shentcis, Reinhardt, Adiv, Wang, Beer, Kurman, Shi, Lynch et~al.}}]{DNS19}
\bibinfo{author}{\bibfnamefont{R.}~\bibnamefont{Dahan}},
  \bibinfo{author}{\bibfnamefont{S.}~\bibnamefont{Nehemia}},
  \bibinfo{author}{\bibfnamefont{M.}~\bibnamefont{Shentcis}},
  \bibinfo{author}{\bibfnamefont{O.}~\bibnamefont{Reinhardt}},
  \bibinfo{author}{\bibfnamefont{Y.}~\bibnamefont{Adiv}},
  \bibinfo{author}{\bibfnamefont{K.}~\bibnamefont{Wang}},
  \bibinfo{author}{\bibfnamefont{O.}~\bibnamefont{Beer}},
  \bibinfo{author}{\bibfnamefont{Y.}~\bibnamefont{Kurman}},
  \bibinfo{author}{\bibfnamefont{X.}~\bibnamefont{Shi}},
  \bibinfo{author}{\bibfnamefont{M.~H.} \bibnamefont{Lynch}},
  \bibnamefont{et~al.}, p. \bibinfo{pages}{arXiv:1909.00757}
  (\bibinfo{year}{2019}).

\bibitem[{\citenamefont{Hachtel et~al.}(2019)\citenamefont{Hachtel, Huang,
  Popovs, Jansone-Popova, Keum, Jakowski, Lovejoy, Dellby, Krivanek, and
  Idrobo}}]{HHP19}
\bibinfo{author}{\bibfnamefont{J.~A.} \bibnamefont{Hachtel}},
  \bibinfo{author}{\bibfnamefont{J.}~\bibnamefont{Huang}},
  \bibinfo{author}{\bibfnamefont{I.}~\bibnamefont{Popovs}},
  \bibinfo{author}{\bibfnamefont{S.}~\bibnamefont{Jansone-Popova}},
  \bibinfo{author}{\bibfnamefont{J.~K.} \bibnamefont{Keum}},
  \bibinfo{author}{\bibfnamefont{J.}~\bibnamefont{Jakowski}},
  \bibinfo{author}{\bibfnamefont{T.~C.} \bibnamefont{Lovejoy}},
  \bibinfo{author}{\bibfnamefont{N.}~\bibnamefont{Dellby}},
  \bibinfo{author}{\bibfnamefont{O.~L.} \bibnamefont{Krivanek}},
  \bibnamefont{and} \bibinfo{author}{\bibfnamefont{J.~C.}
  \bibnamefont{Idrobo}}, \bibinfo{journal}{Science}
  \textbf{\bibinfo{volume}{363}}, \bibinfo{pages}{525} (\bibinfo{year}{2019}).

\bibitem[{\citenamefont{Polman et~al.}(2019)\citenamefont{Polman, Kociak, and
  {Garc\'{\i}a de Abajo}}}]{paper338}
\bibinfo{author}{\bibfnamefont{A.}~\bibnamefont{Polman}},
  \bibinfo{author}{\bibfnamefont{M.}~\bibnamefont{Kociak}}, \bibnamefont{and}
  \bibinfo{author}{\bibfnamefont{F.~J.} \bibnamefont{{Garc\'{\i}a de Abajo}}},
  \bibinfo{journal}{Nat.\ Mater.} \textbf{\bibinfo{volume}{18}},
  \bibinfo{pages}{1158} (\bibinfo{year}{2019}).

\bibitem[{\citenamefont{{Garc\'{\i}a de Abajo}}(2010)}]{paper149}
\bibinfo{author}{\bibfnamefont{F.~J.} \bibnamefont{{Garc\'{\i}a de Abajo}}},
  \bibinfo{journal}{Rev.\ Mod.\ Phys.} \textbf{\bibinfo{volume}{82}},
  \bibinfo{pages}{209} (\bibinfo{year}{2010}).

\bibitem[{\citenamefont{Rossouw and Botton}(2013)}]{RB13}
\bibinfo{author}{\bibfnamefont{D.}~\bibnamefont{Rossouw}} \bibnamefont{and}
  \bibinfo{author}{\bibfnamefont{G.~A.} \bibnamefont{Botton}},
  \bibinfo{journal}{Phys.\ Rev.\ Lett.} \textbf{\bibinfo{volume}{110}},
  \bibinfo{pages}{066801} (\bibinfo{year}{2013}).

\bibitem[{\citenamefont{Kociak and Stephan}(2014)}]{KS14}
\bibinfo{author}{\bibfnamefont{M.}~\bibnamefont{Kociak}} \bibnamefont{and}
  \bibinfo{author}{\bibfnamefont{O.}~\bibnamefont{Stephan}},
  \bibinfo{journal}{Chem.\ Soc.\ Rev.} \textbf{\bibinfo{volume}{43}},
  \bibinfo{pages}{3865} (\bibinfo{year}{2014}).

\bibitem[{\citenamefont{Anton~H\"orl and Hohenester}(2015)}]{HTH15}
\bibinfo{author}{\bibfnamefont{A.~T.} \bibnamefont{Anton~H\"orl}}
  \bibnamefont{and}
  \bibinfo{author}{\bibfnamefont{U.}~\bibnamefont{Hohenester}},
  \bibinfo{journal}{ACS\ Photon.} \textbf{\bibinfo{volume}{2}},
  \bibinfo{pages}{1429} (\bibinfo{year}{2015}).

\bibitem[{\citenamefont{Guzzinati et~al.}(2017)\citenamefont{Guzzinati, Beche,
  Lourenco-Martins, Martin, Kociak, and Verbeeck}}]{GBL17}
\bibinfo{author}{\bibfnamefont{G.}~\bibnamefont{Guzzinati}},
  \bibinfo{author}{\bibfnamefont{A.}~\bibnamefont{Beche}},
  \bibinfo{author}{\bibfnamefont{H.}~\bibnamefont{Lourenco-Martins}},
  \bibinfo{author}{\bibfnamefont{J.}~\bibnamefont{Martin}},
  \bibinfo{author}{\bibfnamefont{M.}~\bibnamefont{Kociak}}, \bibnamefont{and}
  \bibinfo{author}{\bibfnamefont{J.}~\bibnamefont{Verbeeck}},
  \bibinfo{journal}{Nat.\ Commun.} \textbf{\bibinfo{volume}{8}},
  \bibinfo{pages}{14999} (\bibinfo{year}{2017}).

\bibitem[{\citenamefont{Krehl et~al.}(2018)\citenamefont{Krehl, Guzzinati,
  Schultz, Potapov, Pohl, Martin, Verbeeck, Fery, B\"uchner, and Lubk}}]{KGS18}
\bibinfo{author}{\bibfnamefont{J.}~\bibnamefont{Krehl}},
  \bibinfo{author}{\bibfnamefont{G.}~\bibnamefont{Guzzinati}},
  \bibinfo{author}{\bibfnamefont{J.}~\bibnamefont{Schultz}},
  \bibinfo{author}{\bibfnamefont{P.}~\bibnamefont{Potapov}},
  \bibinfo{author}{\bibfnamefont{D.}~\bibnamefont{Pohl}},
  \bibinfo{author}{\bibfnamefont{J.}~\bibnamefont{Martin}},
  \bibinfo{author}{\bibfnamefont{J.}~\bibnamefont{Verbeeck}},
  \bibinfo{author}{\bibfnamefont{A.}~\bibnamefont{Fery}},
  \bibinfo{author}{\bibfnamefont{B.}~\bibnamefont{B\"uchner}},
  \bibnamefont{and} \bibinfo{author}{\bibfnamefont{A.}~\bibnamefont{Lubk}},
  \bibinfo{journal}{Nat.\ Commun.} \textbf{\bibinfo{volume}{9}},
  \bibinfo{pages}{4207} (\bibinfo{year}{2018}).

\bibitem[{\citenamefont{Tizei et~al.}(2015)\citenamefont{Tizei, Lin, Mukai,
  Sawada, Lu, Li, Kimoto, and Suenaga}}]{TLM15}
\bibinfo{author}{\bibfnamefont{L.~H.~G.} \bibnamefont{Tizei}},
  \bibinfo{author}{\bibfnamefont{Y.-C.} \bibnamefont{Lin}},
  \bibinfo{author}{\bibfnamefont{M.}~\bibnamefont{Mukai}},
  \bibinfo{author}{\bibfnamefont{H.}~\bibnamefont{Sawada}},
  \bibinfo{author}{\bibfnamefont{A.-Y.} \bibnamefont{Lu}},
  \bibinfo{author}{\bibfnamefont{L.-J.} \bibnamefont{Li}},
  \bibinfo{author}{\bibfnamefont{K.}~\bibnamefont{Kimoto}}, \bibnamefont{and}
  \bibinfo{author}{\bibfnamefont{K.}~\bibnamefont{Suenaga}},
  \bibinfo{journal}{Phys.\ Rev.\ Lett.} \textbf{\bibinfo{volume}{114}},
  \bibinfo{pages}{107601} (\bibinfo{year}{2015}).

\bibitem[{\citenamefont{Senga et~al.}(2019)\citenamefont{Senga, Suenaga,
  Barone, Morishita, Mauri, and Pichler}}]{SSB19}
\bibinfo{author}{\bibfnamefont{R.}~\bibnamefont{Senga}},
  \bibinfo{author}{\bibfnamefont{K.}~\bibnamefont{Suenaga}},
  \bibinfo{author}{\bibfnamefont{P.}~\bibnamefont{Barone}},
  \bibinfo{author}{\bibfnamefont{S.}~\bibnamefont{Morishita}},
  \bibinfo{author}{\bibfnamefont{F.}~\bibnamefont{Mauri}}, \bibnamefont{and}
  \bibinfo{author}{\bibfnamefont{T.}~\bibnamefont{Pichler}},
  \bibinfo{journal}{Nature} pp. \bibinfo{pages}{247--250}
  (\bibinfo{year}{2019}).

\bibitem[{\citenamefont{Rez et~al.}(2016)\citenamefont{Rez, Aoki, March, Gur,
  Krivanek, Dellby, Lovejoy, Wolf, and Cohen}}]{RAM16}
\bibinfo{author}{\bibfnamefont{P.}~\bibnamefont{Rez}},
  \bibinfo{author}{\bibfnamefont{T.}~\bibnamefont{Aoki}},
  \bibinfo{author}{\bibfnamefont{K.}~\bibnamefont{March}},
  \bibinfo{author}{\bibfnamefont{D.}~\bibnamefont{Gur}},
  \bibinfo{author}{\bibfnamefont{O.~L.} \bibnamefont{Krivanek}},
  \bibinfo{author}{\bibfnamefont{N.}~\bibnamefont{Dellby}},
  \bibinfo{author}{\bibfnamefont{T.~C.} \bibnamefont{Lovejoy}},
  \bibinfo{author}{\bibfnamefont{S.~G.} \bibnamefont{Wolf}}, \bibnamefont{and}
  \bibinfo{author}{\bibfnamefont{H.}~\bibnamefont{Cohen}},
  \bibinfo{journal}{Nat.\ Commun.} \textbf{\bibinfo{volume}{7}},
  \bibinfo{pages}{10945} (\bibinfo{year}{2016}).

\bibitem[{\citenamefont{Morimoto and Baum}(2017)}]{MB17}
\bibinfo{author}{\bibfnamefont{Y.}~\bibnamefont{Morimoto}} \bibnamefont{and}
  \bibinfo{author}{\bibfnamefont{P.}~\bibnamefont{Baum}},
  \bibinfo{journal}{Nat. Phys.} \textbf{\bibinfo{volume}{14}},
  \bibinfo{pages}{252} (\bibinfo{year}{2017}).

\bibitem[{\citenamefont{Kfir}(2019)}]{K19}
\bibinfo{author}{\bibfnamefont{O.}~\bibnamefont{Kfir}},
  \bibinfo{journal}{Phys.\ Rev.\ Lett.} \textbf{\bibinfo{volume}{123}},
  \bibinfo{pages}{103602} (\bibinfo{year}{2019}).

\bibitem[{\citenamefont{Di~Giulio et~al.}(2019)\citenamefont{Di~Giulio, Kociak,
  and Garc\'{i}a~de Abajo}}]{paper3xy}
\bibinfo{author}{\bibfnamefont{V.}~\bibnamefont{Di~Giulio}},
  \bibinfo{author}{\bibfnamefont{M.}~\bibnamefont{Kociak}}, \bibnamefont{and}
  \bibinfo{author}{\bibfnamefont{F.~J.} \bibnamefont{Garc\'{i}a~de Abajo}},
  \bibinfo{journal}{arXiv} \textbf{\bibinfo{volume}{0}},
  \bibinfo{pages}{1905.06887v4} (\bibinfo{year}{2019}).

\bibitem[{\citenamefont{Reinhardt et~al.}(2019)\citenamefont{Reinhardt, Mechel,
  Lynch, and Kaminer}}]{RML19}
\bibinfo{author}{\bibfnamefont{O.}~\bibnamefont{Reinhardt}},
  \bibinfo{author}{\bibfnamefont{C.}~\bibnamefont{Mechel}},
  \bibinfo{author}{\bibfnamefont{M.}~\bibnamefont{Lynch}}, \bibnamefont{and}
  \bibinfo{author}{\bibfnamefont{I.}~\bibnamefont{Kaminer}}, p.
  \bibinfo{pages}{arXiv:1907.10281} (\bibinfo{year}{2019}).

\bibitem[{\citenamefont{Kauranen and Zayats}(2012)}]{KZ12}
\bibinfo{author}{\bibfnamefont{M.}~\bibnamefont{Kauranen}} \bibnamefont{and}
  \bibinfo{author}{\bibfnamefont{A.~V.} \bibnamefont{Zayats}},
  \bibinfo{journal}{Nat.\ Photon.} \textbf{\bibinfo{volume}{6}},
  \bibinfo{pages}{737} (\bibinfo{year}{2012}).

\bibitem[{\citenamefont{Panoiu et~al.}(2018)\citenamefont{Panoiu, Sha, Lei, and
  Li}}]{PSL18}
\bibinfo{author}{\bibfnamefont{N.~C.} \bibnamefont{Panoiu}},
  \bibinfo{author}{\bibfnamefont{W.~E.~I.} \bibnamefont{Sha}},
  \bibinfo{author}{\bibfnamefont{D.~Y.} \bibnamefont{Lei}}, \bibnamefont{and}
  \bibinfo{author}{\bibfnamefont{G.-C.} \bibnamefont{Li}},
  \bibinfo{journal}{J.\ Opt.} \textbf{\bibinfo{volume}{20}},
  \bibinfo{pages}{083001} (\bibinfo{year}{2018}).

\bibitem[{\citenamefont{Butet et~al.}(2015)\citenamefont{Butet, Brevet, and
  Martin}}]{BBM15}
\bibinfo{author}{\bibfnamefont{J.}~\bibnamefont{Butet}},
  \bibinfo{author}{\bibfnamefont{P.-F.} \bibnamefont{Brevet}},
  \bibnamefont{and} \bibinfo{author}{\bibfnamefont{O.~J.~F.}
  \bibnamefont{Martin}}, \bibinfo{journal}{ACS\ Nano}
  \textbf{\bibinfo{volume}{9}}, \bibinfo{pages}{10545} (\bibinfo{year}{2015}).

\bibitem[{\citenamefont{Bozhevolnyi et~al.}(1998)\citenamefont{Bozhevolnyi,
  Pedersen, Skettrup, Zhang, and Belmonte}}]{BPS98}
\bibinfo{author}{\bibfnamefont{S.~I.} \bibnamefont{Bozhevolnyi}},
  \bibinfo{author}{\bibfnamefont{K.}~\bibnamefont{Pedersen}},
  \bibinfo{author}{\bibfnamefont{T.}~\bibnamefont{Skettrup}},
  \bibinfo{author}{\bibfnamefont{X.}~\bibnamefont{Zhang}}, \bibnamefont{and}
  \bibinfo{author}{\bibfnamefont{M.}~\bibnamefont{Belmonte}},
  \bibinfo{journal}{Opt.\ Commun.} \textbf{\bibinfo{volume}{152}},
  \bibinfo{pages}{221} (\bibinfo{year}{1998}).

\bibitem[{\citenamefont{Boyd}(2008)}]{B08_3}
\bibinfo{author}{\bibfnamefont{R.~W.} \bibnamefont{Boyd}},
  \emph{\bibinfo{title}{Nonlinear optics}} (\bibinfo{publisher}{Academic
  Press}, \bibinfo{address}{Amsterdam}, \bibinfo{year}{2008}),
  \bibinfo{edition}{3rd} ed.

\bibitem[{\citenamefont{Zayats and Sandoghdar}(2000)}]{ZS00}
\bibinfo{author}{\bibfnamefont{A.~V.} \bibnamefont{Zayats}} \bibnamefont{and}
  \bibinfo{author}{\bibfnamefont{V.}~\bibnamefont{Sandoghdar}},
  \bibinfo{journal}{Opt.\ Commun.} \textbf{\bibinfo{volume}{178}},
  \bibinfo{pages}{245} (\bibinfo{year}{2000}).

\bibitem[{\citenamefont{Bouhelier et~al.}(2003)\citenamefont{Bouhelier,
  Beversluis, Hartschuh, and Novotny}}]{BBH03}
\bibinfo{author}{\bibfnamefont{A.}~\bibnamefont{Bouhelier}},
  \bibinfo{author}{\bibfnamefont{M.}~\bibnamefont{Beversluis}},
  \bibinfo{author}{\bibfnamefont{A.}~\bibnamefont{Hartschuh}},
  \bibnamefont{and} \bibinfo{author}{\bibfnamefont{L.}~\bibnamefont{Novotny}},
  \bibinfo{journal}{Phys.\ Rev.\ Lett.} \textbf{\bibinfo{volume}{90}},
  \bibinfo{pages}{013903} (\bibinfo{year}{2003}).

\bibitem[{\citenamefont{Zavelani-Rossi
  et~al.}(2008)\citenamefont{Zavelani-Rossi, Celebrano, Biagioni, Polli,
  Finazzi, Duo, Cerullo, Labardi, Allegrini, Grand et~al.}}]{ZCB08}
\bibinfo{author}{\bibfnamefont{M.}~\bibnamefont{Zavelani-Rossi}},
  \bibinfo{author}{\bibfnamefont{M.}~\bibnamefont{Celebrano}},
  \bibinfo{author}{\bibfnamefont{P.}~\bibnamefont{Biagioni}},
  \bibinfo{author}{\bibfnamefont{D.}~\bibnamefont{Polli}},
  \bibinfo{author}{\bibfnamefont{M.}~\bibnamefont{Finazzi}},
  \bibinfo{author}{\bibfnamefont{L.}~\bibnamefont{Duo}},
  \bibinfo{author}{\bibfnamefont{G.}~\bibnamefont{Cerullo}},
  \bibinfo{author}{\bibfnamefont{M.}~\bibnamefont{Labardi}},
  \bibinfo{author}{\bibfnamefont{M.}~\bibnamefont{Allegrini}},
  \bibinfo{author}{\bibfnamefont{J.}~\bibnamefont{Grand}},
  \bibnamefont{et~al.}, \bibinfo{journal}{Appl.\ Phys.\ Lett.}
  \textbf{\bibinfo{volume}{92}}, \bibinfo{pages}{093119}
  (\bibinfo{year}{2008}).

\bibitem[{\citenamefont{Neacsu et~al.}(2009)\citenamefont{Neacsu, {van Aken},
  Fiebig, and Raschke}}]{NVF09}
\bibinfo{author}{\bibfnamefont{C.~C.} \bibnamefont{Neacsu}},
  \bibinfo{author}{\bibfnamefont{B.~B.} \bibnamefont{{van Aken}}},
  \bibinfo{author}{\bibfnamefont{M.}~\bibnamefont{Fiebig}}, \bibnamefont{and}
  \bibinfo{author}{\bibfnamefont{M.~B.} \bibnamefont{Raschke}},
  \bibinfo{journal}{Phys.\ Rev.\ B} \textbf{\bibinfo{volume}{79}},
  \bibinfo{pages}{100107(R)} (\bibinfo{year}{2009}).

\bibitem[{\citenamefont{Metzger et~al.}(2017)\citenamefont{Metzger, Hentschel,
  and Giessen}}]{MHG17}
\bibinfo{author}{\bibfnamefont{B.}~\bibnamefont{Metzger}},
  \bibinfo{author}{\bibfnamefont{M.}~\bibnamefont{Hentschel}},
  \bibnamefont{and} \bibinfo{author}{\bibfnamefont{H.}~\bibnamefont{Giessen}},
  \bibinfo{journal}{Nano\ Lett.} \textbf{\bibinfo{volume}{17}},
  \bibinfo{pages}{1931} (\bibinfo{year}{2017}).

\bibitem[{\citenamefont{Howie}(1999)}]{H99}
\bibinfo{author}{\bibfnamefont{A.}~\bibnamefont{Howie}},
  \bibinfo{journal}{Inst. Phys. Conf. Ser.} \textbf{\bibinfo{volume}{161}},
  \bibinfo{pages}{311} (\bibinfo{year}{1999}).

\bibitem[{\citenamefont{{Garc\'{\i}a de Abajo} and Kociak}(2008)}]{paper114}
\bibinfo{author}{\bibfnamefont{F.~J.} \bibnamefont{{Garc\'{\i}a de Abajo}}}
  \bibnamefont{and} \bibinfo{author}{\bibfnamefont{M.}~\bibnamefont{Kociak}},
  \bibinfo{journal}{New\ J.\ Phys.} \textbf{\bibinfo{volume}{10}},
  \bibinfo{pages}{073035} (\bibinfo{year}{2008}).

\bibitem[{\citenamefont{{Garc\'{\i}a de Abajo}
  et~al.}(2016)\citenamefont{{Garc\'{\i}a de Abajo}, Barwick, and
  Carbone}}]{paper272}
\bibinfo{author}{\bibfnamefont{F.~J.} \bibnamefont{{Garc\'{\i}a de Abajo}}},
  \bibinfo{author}{\bibfnamefont{B.}~\bibnamefont{Barwick}}, \bibnamefont{and}
  \bibinfo{author}{\bibfnamefont{F.}~\bibnamefont{Carbone}},
  \bibinfo{journal}{Phys.\ Rev.\ B} \textbf{\bibinfo{volume}{94}},
  \bibinfo{pages}{041404(R)} (\bibinfo{year}{2016}).

\bibitem[{\citenamefont{Bloembergen et~al.}(1968)\citenamefont{Bloembergen,
  Chang, Jha, and Lee}}]{BCJ1968}
\bibinfo{author}{\bibfnamefont{N.}~\bibnamefont{Bloembergen}},
  \bibinfo{author}{\bibfnamefont{R.~K.} \bibnamefont{Chang}},
  \bibinfo{author}{\bibfnamefont{S.~S.} \bibnamefont{Jha}}, \bibnamefont{and}
  \bibinfo{author}{\bibfnamefont{C.~H.} \bibnamefont{Lee}},
  \bibinfo{journal}{Phys.\ Rev.} \textbf{\bibinfo{volume}{174}},
  \bibinfo{pages}{813} (\bibinfo{year}{1968}).

\bibitem[{\citenamefont{Simon et~al.}(1974)\citenamefont{Simon, Mitchell, and
  Watson}}]{SMW1974}
\bibinfo{author}{\bibfnamefont{H.~J.} \bibnamefont{Simon}},
  \bibinfo{author}{\bibfnamefont{D.~E.} \bibnamefont{Mitchell}},
  \bibnamefont{and} \bibinfo{author}{\bibfnamefont{J.~G.}
  \bibnamefont{Watson}}, \bibinfo{journal}{Phys.\ Rev.\ Lett.}
  \textbf{\bibinfo{volume}{33}}, \bibinfo{pages}{1531} (\bibinfo{year}{1974}).

\bibitem[{\citenamefont{Sipe et~al.}(1980)\citenamefont{Sipe, So, Fukui, and
  Stegeman}}]{SSF1980}
\bibinfo{author}{\bibfnamefont{J.~E.} \bibnamefont{Sipe}},
  \bibinfo{author}{\bibfnamefont{V.~C.~Y.} \bibnamefont{So}},
  \bibinfo{author}{\bibfnamefont{M.}~\bibnamefont{Fukui}}, \bibnamefont{and}
  \bibinfo{author}{\bibfnamefont{G.~I.} \bibnamefont{Stegeman}},
  \bibinfo{journal}{Phys.\ Rev.\ B} \textbf{\bibinfo{volume}{21}},
  \bibinfo{pages}{4389} (\bibinfo{year}{1980}).

\bibitem[{\citenamefont{Galanty et~al.}(2018)\citenamefont{Galanty, Shavit,
  Weissman, Aharon, Gachet, Segal, and Salomon}}]{GSW18}
\bibinfo{author}{\bibfnamefont{M.}~\bibnamefont{Galanty}},
  \bibinfo{author}{\bibfnamefont{O.}~\bibnamefont{Shavit}},
  \bibinfo{author}{\bibfnamefont{A.}~\bibnamefont{Weissman}},
  \bibinfo{author}{\bibfnamefont{H.}~\bibnamefont{Aharon}},
  \bibinfo{author}{\bibfnamefont{D.}~\bibnamefont{Gachet}},
  \bibinfo{author}{\bibfnamefont{E.}~\bibnamefont{Segal}}, \bibnamefont{and}
  \bibinfo{author}{\bibfnamefont{A.}~\bibnamefont{Salomon}},
  \bibinfo{journal}{Light\ Sci.\ Appl.} \textbf{\bibinfo{volume}{7}},
  \bibinfo{pages}{49} (\bibinfo{year}{2018}).

\bibitem[{\citenamefont{Bachelier et~al.}(2010)\citenamefont{Bachelier, Butet,
  Russier-Antoine, Jonin, Benichou, and Brevet}}]{BBR10_2}
\bibinfo{author}{\bibfnamefont{G.}~\bibnamefont{Bachelier}},
  \bibinfo{author}{\bibfnamefont{J.}~\bibnamefont{Butet}},
  \bibinfo{author}{\bibfnamefont{I.}~\bibnamefont{Russier-Antoine}},
  \bibinfo{author}{\bibfnamefont{C.}~\bibnamefont{Jonin}},
  \bibinfo{author}{\bibfnamefont{E.}~\bibnamefont{Benichou}}, \bibnamefont{and}
  \bibinfo{author}{\bibfnamefont{P.-F.} \bibnamefont{Brevet}},
  \bibinfo{journal}{Phys.\ Rev.\ B} \textbf{\bibinfo{volume}{82}},
  \bibinfo{pages}{235403} (\bibinfo{year}{2010}).

\bibitem[{\citenamefont{Krause et~al.}(2004)\citenamefont{Krause, Teplin, and
  Rogers}}]{KTR04}
\bibinfo{author}{\bibfnamefont{D.}~\bibnamefont{Krause}},
  \bibinfo{author}{\bibfnamefont{C.~W.} \bibnamefont{Teplin}},
  \bibnamefont{and} \bibinfo{author}{\bibfnamefont{C.~T.}
  \bibnamefont{Rogers}}, \bibinfo{journal}{J.\ Appl.\ Phys.}
  \textbf{\bibinfo{volume}{96}}, \bibinfo{pages}{3626} (\bibinfo{year}{2004}).

\bibitem[{\citenamefont{Wang et~al.}(2009)\citenamefont{Wang, Rodr\'{\i}guez,
  Albers, Ahorinta, Sipe, and Kauranen}}]{WRA09}
\bibinfo{author}{\bibfnamefont{F.~X.} \bibnamefont{Wang}},
  \bibinfo{author}{\bibfnamefont{F.~J.} \bibnamefont{Rodr\'{\i}guez}},
  \bibinfo{author}{\bibfnamefont{W.~M.} \bibnamefont{Albers}},
  \bibinfo{author}{\bibfnamefont{R.}~\bibnamefont{Ahorinta}},
  \bibinfo{author}{\bibfnamefont{J.~E.} \bibnamefont{Sipe}}, \bibnamefont{and}
  \bibinfo{author}{\bibfnamefont{M.}~\bibnamefont{Kauranen}},
  \bibinfo{journal}{Phys.\ Rev.\ B} \textbf{\bibinfo{volume}{80}},
  \bibinfo{pages}{233402} (\bibinfo{year}{2009}).

\bibitem[{\citenamefont{Timbrell et~al.}(2018)\citenamefont{Timbrell, You,
  Kivshar, and Panoiu}}]{TYK18}
\bibinfo{author}{\bibfnamefont{D.}~\bibnamefont{Timbrell}},
  \bibinfo{author}{\bibfnamefont{J.~W.} \bibnamefont{You}},
  \bibinfo{author}{\bibfnamefont{Y.~S.} \bibnamefont{Kivshar}},
  \bibnamefont{and} \bibinfo{author}{\bibfnamefont{N.~C.}
  \bibnamefont{Panoiu}}, \bibinfo{journal}{Sci.\ Rep.}
  \textbf{\bibinfo{volume}{8}}, \bibinfo{pages}{3586} (\bibinfo{year}{2018}).

\bibitem[{\citenamefont{Johnson and Christy}(1972)}]{JC1972}
\bibinfo{author}{\bibfnamefont{P.~B.} \bibnamefont{Johnson}} \bibnamefont{and}
  \bibinfo{author}{\bibfnamefont{R.~W.} \bibnamefont{Christy}},
  \bibinfo{journal}{Phys.\ Rev.\ B} \textbf{\bibinfo{volume}{6}},
  \bibinfo{pages}{4370} (\bibinfo{year}{1972}).

\bibitem[{\citenamefont{Dadap et~al.}(1999)\citenamefont{Dadap, Shan,
  Eisenthal, and Heinz}}]{DSE1999}
\bibinfo{author}{\bibfnamefont{J.~I.} \bibnamefont{Dadap}},
  \bibinfo{author}{\bibfnamefont{J.}~\bibnamefont{Shan}},
  \bibinfo{author}{\bibfnamefont{K.~B.} \bibnamefont{Eisenthal}},
  \bibnamefont{and} \bibinfo{author}{\bibfnamefont{T.~F.} \bibnamefont{Heinz}},
  \bibinfo{journal}{Phys.\ Rev.\ Lett.} \textbf{\bibinfo{volume}{83}},
  \bibinfo{pages}{4045} (\bibinfo{year}{1999}).

\bibitem[{\citenamefont{Dadap et~al.}(2004)\citenamefont{Dadap, Shan, and
  Heinz}}]{DSH04}
\bibinfo{author}{\bibfnamefont{J.~I.} \bibnamefont{Dadap}},
  \bibinfo{author}{\bibfnamefont{J.}~\bibnamefont{Shan}}, \bibnamefont{and}
  \bibinfo{author}{\bibfnamefont{T.~F.} \bibnamefont{Heinz}},
  \bibinfo{journal}{J.\ Opt.\ Soc.\ Am.\ B} \textbf{\bibinfo{volume}{21}},
  \bibinfo{pages}{1328} (\bibinfo{year}{2004}).

\bibitem[{\citenamefont{Russier-Antoine
  et~al.}(2007)\citenamefont{Russier-Antoine, Benichou, Bachelier, Jonin, and
  Brevet}}]{RBB07}
\bibinfo{author}{\bibfnamefont{I.}~\bibnamefont{Russier-Antoine}},
  \bibinfo{author}{\bibfnamefont{E.}~\bibnamefont{Benichou}},
  \bibinfo{author}{\bibfnamefont{G.}~\bibnamefont{Bachelier}},
  \bibinfo{author}{\bibfnamefont{C.}~\bibnamefont{Jonin}}, \bibnamefont{and}
  \bibinfo{author}{\bibfnamefont{P.~F.} \bibnamefont{Brevet}},
  \bibinfo{journal}{J.\ Phys.\ Chem.\ C} \textbf{\bibinfo{volume}{111}},
  \bibinfo{pages}{9044} (\bibinfo{year}{2007}).

\bibitem[{\citenamefont{Jaynes and Cummings}(1963)}]{JC1963}
\bibinfo{author}{\bibfnamefont{E.}~\bibnamefont{Jaynes}} \bibnamefont{and}
  \bibinfo{author}{\bibfnamefont{F.}~\bibnamefont{Cummings}},
  \bibinfo{journal}{Proc.\ IEEE} \textbf{\bibinfo{volume}{51}},
  \bibinfo{pages}{89} (\bibinfo{year}{1963}).

\bibitem[{\citenamefont{Dattoli et~al.}(1996)\citenamefont{Dattoli, Chiccoli,
  Lorenzutta, Maino, Richetta, and Torre}}]{DTL1996}
\bibinfo{author}{\bibfnamefont{G.}~\bibnamefont{Dattoli}},
  \bibinfo{author}{\bibfnamefont{C.}~\bibnamefont{Chiccoli}},
  \bibinfo{author}{\bibfnamefont{S.}~\bibnamefont{Lorenzutta}},
  \bibinfo{author}{\bibfnamefont{G.}~\bibnamefont{Maino}},
  \bibinfo{author}{\bibfnamefont{M.}~\bibnamefont{Richetta}}, \bibnamefont{and}
  \bibinfo{author}{\bibfnamefont{A.}~\bibnamefont{Torre}},
  \bibinfo{journal}{Radiat.\ Phys.\ Chem.} \textbf{\bibinfo{volume}{47}},
  \bibinfo{pages}{183} (\bibinfo{year}{1996}).

\bibitem[{\citenamefont{Jackson}(1999)}]{J99}
\bibinfo{author}{\bibfnamefont{J.~D.} \bibnamefont{Jackson}},
  \emph{\bibinfo{title}{Classical Electrodynamics}}
  (\bibinfo{publisher}{Wiley}, \bibinfo{address}{New York},
  \bibinfo{year}{1999}).

\bibitem[{\citenamefont{Messiah}(1966)}]{M1966}
\bibinfo{author}{\bibfnamefont{A.}~\bibnamefont{Messiah}},
  \emph{\bibinfo{title}{Quantum Mechanics}}
  (\bibinfo{publisher}{North-Holland}, \bibinfo{address}{New York},
  \bibinfo{year}{1966}).

\bibitem[{\citenamefont{Gradshteyn and Ryzhik}(2007)}]{GR1980}
\bibinfo{author}{\bibfnamefont{I.~S.} \bibnamefont{Gradshteyn}}
  \bibnamefont{and} \bibinfo{author}{\bibfnamefont{I.~M.}
  \bibnamefont{Ryzhik}}, \emph{\bibinfo{title}{Table of Integrals, Series, and
  Products}} (\bibinfo{publisher}{Academic Press}, \bibinfo{address}{London},
  \bibinfo{year}{2007}).

\end{thebibliography}

\end{document}